# A Novel Context driven Critical Integrative Levels (CIL) Approach: Advancing Human-Centric and Integrative Lighting Asset Management in Public Libraries with Practical Thresholds

Jing Lin[1,2*], Nina Mylly[3], Per Olof Hedekvist[3], Jingchun Shen[4,5]

1. Division of Operation and Maintenance, Luleå University of Technology, 97187, Luleå, Sweden
2. Division of Product Realization, Mälardalen University, 63220, Eskilstuna, Sweden
3. Division of Measurement Techniques, Research Institutes of Sweden, 50462, Borås, Sweden
4. Division of Construction Technology, Dalarna University, 79188, Falun, Sweden
5. Sustainable Energy Research Centre, Dalarna University, 79188, Falun, Sweden

**Abstract:** Addressing the challenge of managing human-centric and integrative lighting in public libraries, especially post-installation, requires balancing complex visual and non-visual light effects. Current standards may overlook varied user needs and environmental contexts, potentially hindering effective post-installation asset management and digitalization efforts. To solve this problem, this paper proposes the context driven Critical Integrative Levels (CIL), a novel approach to lighting asset management in public libraries that aligns with the transformative vision of human-centric and integrative lighting. This approach encompasses not only the visual aspects of lighting performance but also prioritizes the physiological and psychological well-being of library users. With the Global Lighting Association (GLA)'s aim to realize human-centric lighting by 2040, this study contributes a proactive methodology that goes beyond traditional asset management concerns of functionality and lifespan to embrace user well-being. Incorporating a newly defined metric, Mean Time of Exposure (MTOE), the approach quantifies user-light interaction, enabling tailored lighting strategies that respond to diverse activities and needs in library spaces. Case studies demonstrate how the CIL matrix can be practically applied, offering significant improvements over conventional methods by focusing on optimized user experiences from both visual impacts and non-visual effects. Results from this study highlight the study's implications for future policy and practice in lighting asset management, anticipating further research to extend the scope and impact of human-centric lighting in line with industry forecasts.

## 1. Introduction

"Human-centric lighting" [1, 2], more broadly recognized as "integrative lighting" [3], is specifically designed to positively influence human physiology and psychology. This encompasses both visual aspects—such as light intensity, flicker, glare, and colour temperature—and non-visual impacts, including biological and emotional advantages like circadian rhythm synchronization and heightened alertness [4, 5]. The Global Lighting Association (GLA) has charted a strategic roadmap toward the realization of human-centric lighting by 2040 [1]. This study, supported by the Swedish Energy Agency, seeks to pioneer advancements in human-centric and integrative lighting asset management, which aims to promote effective lighting and to help the users get the right light at the right time to support various activities, with public libraries serving as the illustrative context.

Contrasting with traditional asset management's focus on functionality, cost-efficiency, and asset longevity, human-centric and integrative asset management centralizes human well-being in its decision-making processes. It seeks to refine technical specifications to bolster human health and wellness [6] [7].

Today, public libraries are transitioning from mere repositories of books to hubs of learning. Indoor lighting quality in such venues doesn't just influence physical health, like vision, but also factors into mental well-being, influencing moods and productivity [8] [9]. A pressing challenge within the domain of public library lighting asset management is how to manage human-centric and integrative lighting assets, particular in post-installation stage, given the increasing requirements and complexity involved when combining the visual and non-visual effects of light [6] [7].

* Corresponding Author: Janet (Jing) Lin, Email: janet.lin@ltu.se; janet.lin@mdu.se

Using illuminance as an example, preliminary investigations within a single public library demonstrate that the health status of a lighting asset can be assessed by comparing real-time measurements against established norms and guidelines [7]. Specifically, the disparity between these real-time measurements and the normative values offers a quantifiable basis for identifying potential health risks (see Table.1 in [7]). Such a methodology not only permits the establishment of anomaly detection alerts but also enhances diagnostic and prognostic efforts, laying the groundwork for developing key performance indicators (KPIs) and enabling effective monitoring and forecasting strategies. This method would further facilitate the digitalization and visualization of lighting asset management through the implementation of digital twins, thereby enhancing resource/energy efficiency and decision-making processes. However, these efforts represent merely the initial phase. The problem lies in the fact that current thresholds in the above preliminary investigations are established solely based on standard recommendations, neglecting the diverse contexts (e.g., ages of the users, various activities, geographic locations, and seasonal changes, etc.) required to fulfil the demands of human-centric and integrative lighting asset management [7]. Even though we would assume that most lighting design are qualified to be human-centric and integrative, this oversight could still lead to inappropriate in operation and maintenance in post-installation stage of lighting assets [6]; and as the process of digitalization and visualization evolves, it will become obstacle of moving forward to realizing the human-centric and integrative asset management.

To address the above problem, this study introduces the concept of Mean Time of Exposure (MTOE) and proposes a novel, context-driven Critical Integrative Levels (CIL) for the management of lighting assets in public libraries. Furthermore, it establishes and delineates context-specific thresholds for crucial lighting parameters, situated within the domain of human-centric and integrative lighting asset management. Through the implementation of the proposed context-driven CIL, decision-making processes are oriented to prioritize human well-being, thereby assisting public libraries in advancing toward the objective of facilitating human centric and integrative lighting. This ensures users receive the appropriate lighting performance at the precise time necessary to support a variety of activities.

Following the introduction, the structure of this paper is arranged to explore and elucidate the intricacies of human-centric and integrative lighting asset management in public libraries, guided by the proposed Critical Integrative Levels (CIL) and the concept of Mean Time of Exposure (MTOE). Chapter 2 lays the groundwork with an identification of key performance parameters and contexts of lighting in public libraries. Chapter 3 delves into the MTOE and CIL, highlighting their integration to refine management practices. Chapter 4 discusses the establishment of context-specific CIL thresholds, enriched with illustrative examples. Chapter 5 transitions to the practical application of the CIL framework through case studies, Chapter 6 discusses the principal findings from our studies, focusing on the degradation of key lighting parameters, nuances of Equivalent Melanopic Lux (EML), comparative measurements across different seasons and libraries, and the intricacies of energy efficiency. Chapter 7 provides the conclusion together with future work. This compact structure aims to offer a comprehensive yet concise exploration of innovative strategies for managing lighting assets in the public libraries, emphasizing human well-being and operational efficiency.

## 2. Background: Lighting Parameters and Contexts in Public Libraries

Chapter 2 provides essential lighting parameters identified on its performance measurement to public libraries and describes the typical lighting contexts these institutions encounter. This sets the stage for a deeper understanding of the contents in Chapter 3 and 4.

### 2.1 Identification of key lighting parameters on its performance measurement

This section identifies key parameters that reflect the performance of lighting, which refer to: illuminance, glare, colour rendering, colour temperature, flicker, chromaticity, photobiological safety, Equivalent Melanopic Lux (EML), energy efficiency, life span, and other factors. More sub parameters are listed in Table 2.1 (also briefly illustrated by Figure 1 in [6]). These parameters provide a comprehensive view of a lighting system's performance and how well it meets the needs and preferences of users. Standards for acceptable levels of most of these parameters are set by organizations like the Illuminating Engineering Society (IES) and the International Electrotechnical Commission (IEC). An overview of those key lighting parameters on its performance measurement is presented:

- Illuminance is a fundamental lighting performance parameter that measures the amount of light incident on a surface. It is expressed in lux (lx), where one lux is equal to one lumen per square meter. This parameter is crucial as it directly relates to how well a task can be visually performed, and it influences comfort and productivity. Related indices to illuminance help in determining the effectiveness and suitability of a lighting system for a given space and task requirements. They are important for assessing visibility, ensuring comfort, and promoting energy efficiency.
- Glare is a significant lighting performance parameter which refers to the discomfort or impairment of vision a person can experience when parts of their visual field are much brighter than the level to which their eyes are adapted. Glare can be detrimental to both comfort and safety, especially in tasks that require precision and focus. Related indices to glare help in assessing and mitigating the negative impacts of glare in various lighting situations. They are important in maintaining visual comfort and ensuring that lighting systems are designed and installed in a way that minimizes glare.
- Colour rendering refers to how accurately a light source reveals the colour s of objects, people, and surroundings in comparison to a natural light source (usually sunlight). The quality of colour rendering can greatly impact the appearance of a space and the activities within it. Related indices to colour rendering help in choosing the appropriate light source for various applications where the appearance of colour is critical, such as retail, art galleries, healthcare facilities, and more.
- Colour temperature is a measure that describes the spectral characteristics of a white light source, particularly the "warm" or "cool" qualities of the light. The colour temperature of a light source is determined by comparing its hue with a theoretical, heated black-body radiator. The temperature at which the heated black-body radiator matches the hue of the light source is the light's colour temperature, measured in degrees Kelvin (K). Related indices to colour temperature help in choosing the appropriate light source for various applications where the colour temperature plays a significant role, such as mood setting, light satisfaction and aesthetic purposes in interior design, film, and photography, and many more.

- Flicker is a rapid change in brightness of a light source over time. It is often caused by fluctuations in the electrical power supply. High levels of flicker can cause discomfort, fatigue, and even seizures in individuals with photosensitive epilepsy. Even at levels that are undetectable to the human eye, flicker can cause headaches and visual impairment. Related indices to flicker help in evaluating and minimizing the flicker in a lighting system, thereby ensuring visual comfort, and minimizing potential health risks.
- Chromaticity describes the specific colour of a light, independent of its luminance. It is usually represented in a two-dimensional chromaticity diagram, where each point corresponds to a specific colour. Related indices to chromaticity help in choosing the appropriate light source for various applications where the exact colour of the light is important, such as film and photography, retail, museums, and galleries, and more.
- Photobiological Safety refers to the potential risk of photochemical damage to the eye caused by electromagnetic radiation from light sources. The primary wavelengths of interest are the visible blue region between 400 nm and 500 nm. While modern LED light sources emit negligible amount of UV (Ultraviolet) or IR (Infrared radiation), the small size combined with substantial blue emission causes a high blue light hazard. The evaluation and classification are described in EN 62471 [10] and is required for luminaires installed in workplaces within EU through EU directive [11].
- Equivalent Melanopic Lux (EML) is a crucial metric for assessing the impact of light on human health, focusing on its influence on circadian rhythms, sleep, and cognitive functions [12] [13]. It reflects the interaction of light with the eye's photoreceptors, particularly the melanopsin-containing intrinsically photosensitive retinal ganglion cells (ipRGCs), which play a vital role in regulating emotions, alertness, and body temperature. Disruptions in ipRGC function due to inappropriate artificial lighting can result in sleep and mood disorders, among other health issues.
- Energy efficiency in lighting refers to the amount of light produced per unit of energy consumed. It is a crucial parameter in the design and operation of lighting systems, as improving energy efficiency can significantly reduce energy costs and environmental impacts. Related indices to energy efficiency help to evaluate and compare the energy efficiency of different lighting systems and to design lighting systems that minimize energy consumption and costs.
- The lifespan of a lighting source refers to the duration of its operation until it reaches a point where it can no longer perform its function adequately. It is a crucial factor when considering the cost, maintenance, and environmental impact of lighting systems. Related indices to lifespan help in planning the maintenance and replacement schedule of a lighting system, as well as in calculating the total cost of ownership, which includes not only the initial cost of the light source but also the costs of energy and maintenance over its operational life.
- Other Factors related to Luminaire: A luminaire, also known as a light fixture, is a complete lighting unit that includes a light source or light sources along with the parts designed to distribute the light, to position and protect the light sources, and to connect the light sources to the power supply. These indices help in selecting the appropriate luminaire for a particular application, considering both the lighting requirements and the energy efficiency.

## 2.2 Lighting contexts encountered in public libraries and their impact

A context-driven approach in human-centric and integrative lighting asset management, particularly within public libraries, is pivotal for addressing the unique spectrum of challenges, user needs, and specific conditions inherent in such settings [7]. This approach acknowledges that libraries are not monolithic; they are dynamic environments influenced by a variety of factors. These factors, as detailed in previous studies and encapsulated [7], fall into several distinct categories: 1) External Environmental Factors, which include the impact of natural light and weather conditions; 2) Interior Design Factors, which pertain to the architectural layout and aesthetic considerations; 3) User Factors, focusing on the demographic diversity and the varying needs of library patrons; 4) Cost Factors, concerning budgetary limitations and economic efficiency; and 5) Regulation Factors, which involve compliance with legal and safety standards.

The human-centric and integrative approach to lighting asset management should acknowledge and cater to the varied needs of these demographic segments. For example, lighting requirements often differ across age groups. In sections of the library such as the magazine and newspaper area, which older adults commonly use, lighting with higher illuminance levels proves more suitable. Similarly, distinct activities necessitate unique lighting conditions. Reading, for instance, demands concentrated, brighter illumination, whereas social gatherings may be enhanced by softer, more diffuse lighting. Additionally, considering the library as a workplace for librarians introduces further lighting considerations. Task-specific lighting must be implemented in areas like the circulation desk and office spaces to reduce eye strain and improve efficiency. Ergonomic lighting design can also support librarians' health and productivity, highlighting the need for a multifaceted approach to lighting that considers the diverse roles a library plays for patrons and staff alike. For libraries situated in high-latitude areas of Sweden (above 55° N), the need for artificial lighting varies significantly with the seasons, attributable to the abundance of daylight in summer and prolonged darkness in winter. Consequently, this necessitates the adoption of diverse operational strategies to accommodate these seasonal variations in lighting requirements.

Understanding and recognizing these contexts is not merely beneficial but foundational to the process of lighting asset management in libraries. By thoroughly grasping the specific situations, settings, or environments, stakeholders can devise and implement a management approach that is not only tailored but also responsive to these diverse influences. This ensures that lighting solutions are not only technically viable but also human-centric, enhancing the user experience by meeting the specific needs dictated by these varied contexts.

In the preliminary framework of lighting asset management, the context-driven approach can offer essential guidance for establishing practical thresholds for the Key Performance Indicators (KPIs) of lighting assets during the "Study" stages [7]. This approach is crucial for addressing the issue that current thresholds, as identified in preliminary investigations, are determined solely based on standard recommendations. By incorporating a context-driven approach, we can ensure that lighting management strategies are more aligned with the actual needs and conditions of library environments, thereby enhancing the effectiveness and relevance of lighting solutions.

In essence, a context-driven approach enables the creation of lighting environments that are adaptable, efficient, and conducive to the multifaceted functions of modern public libraries, thereby underscoring its importance in the realm of library management. Additionally, our study adopts a dynamic approach to setting context-driven thresholds, which are not established as fixed, singular values but rather as a flexible matrix. This matrix is designed to adapt and meet the evolving requirements of human-centric and integrative lighting asset management. For more detailed information, please refer to Chapter 3.

## 3. Novel Thresholds: Mean Time of Exposure (MTOE) and the Multidimensional Critical Integrative Levels (CIL)

Chapter 3 delves into the essence of human-centric and integrative lighting asset management, which prioritizes human well-being at the core of its decision-making process. This chapter introduces the novel concept of Mean Time of Exposure (MTOE) and present a novel, context-driven Critical Integrative Levels (CIL) approach for the effective management of lighting assets in public libraries. This includes an overview of assumptions and illustrative examples. The emphasis is placed on how the integration of these concepts can enhance management practices, ensuring that lighting solutions are not only efficient but also aligned with the specific needs and well-being of library users.

### 3.1 Mean Time of Exposure (MTOE) and Critical Integrative Levels (CIL)

Mean Time of Exposure (MTOE) is a concept designed to quantify the duration of human exposure to specified lighting conditions within a particular environment. It measures the average time individuals spend under these lighting conditions, reflecting the cumulative impact of lighting on human well-being and comfort. MTOE is a critical indicator in evaluating the suitability and effectiveness of lighting environments, particularly in spaces (such as public library) where visual tasks are performed or where lighting significantly influences mood, productivity, and circadian rhythms. MTOE, measured in minutes, is posited to correlate directly with the significance of a given environment to human well-being; that is, a larger MTOE indicates a greater importance of the environment to users' well-being. By assessing MTOE, stakeholders can make informed decisions to optimize lighting designs and configurations, ensuring they meet human-centric needs and enhance the overall user experience.

Critical Integrative Levels (CIL) signify the importance of a lighting environment, with a higher CIL denoting greater significance to human well-being, encompassing both physiological and psychological aspects. CIL determination integrates Mean Time of Exposure (MTOE) with various contextual factors. In this research, the innovative concept of CIL is depicted through a matrix that considers the user's age and MTOE within a public library setting. This study establishes several assumptions to elucidate the novel concept of CIL:

**Assumption 1**: The criticality of an environment is categorized into a higher Critical Integrative Level (CIL) if the MTOE is 15 minutes or more. Conversely, an MTOE of less than 15 minutes categorizes the environment into a lower Critical Integrative Level.

- Higher CIL examples include settings such as reading sofas/tables, learning tables, and counters.
- Lower CIL examples encompass areas like bookshelves, digital inquiry tables, and corridors.

**Assumption 2**: The well-being of users of varying ages is considered differently, affecting lighting requirements. Environments predominantly used by individuals younger than 12 years or older than 65 years are assigned a higher CIL, reflecting the heightened importance of lighting for these age groups.

### 3.2 Establishing Thresholds via Context-Driven Critical Integrative Levels (CIL)

This section details a step-by-step methodology for establishing thresholds using the context-driven CIL approach. The procedure of establishing the thresholds for a single lighting parameter is shown in below Figure. Note: The target value (T) corresponds to the specifications set by the lighting designer, which are confirmed prior to the installation of the lighting assets. Typically, target values are set to achieve higher levels of lighting performance, whereas standard values define the basic acceptable levels.

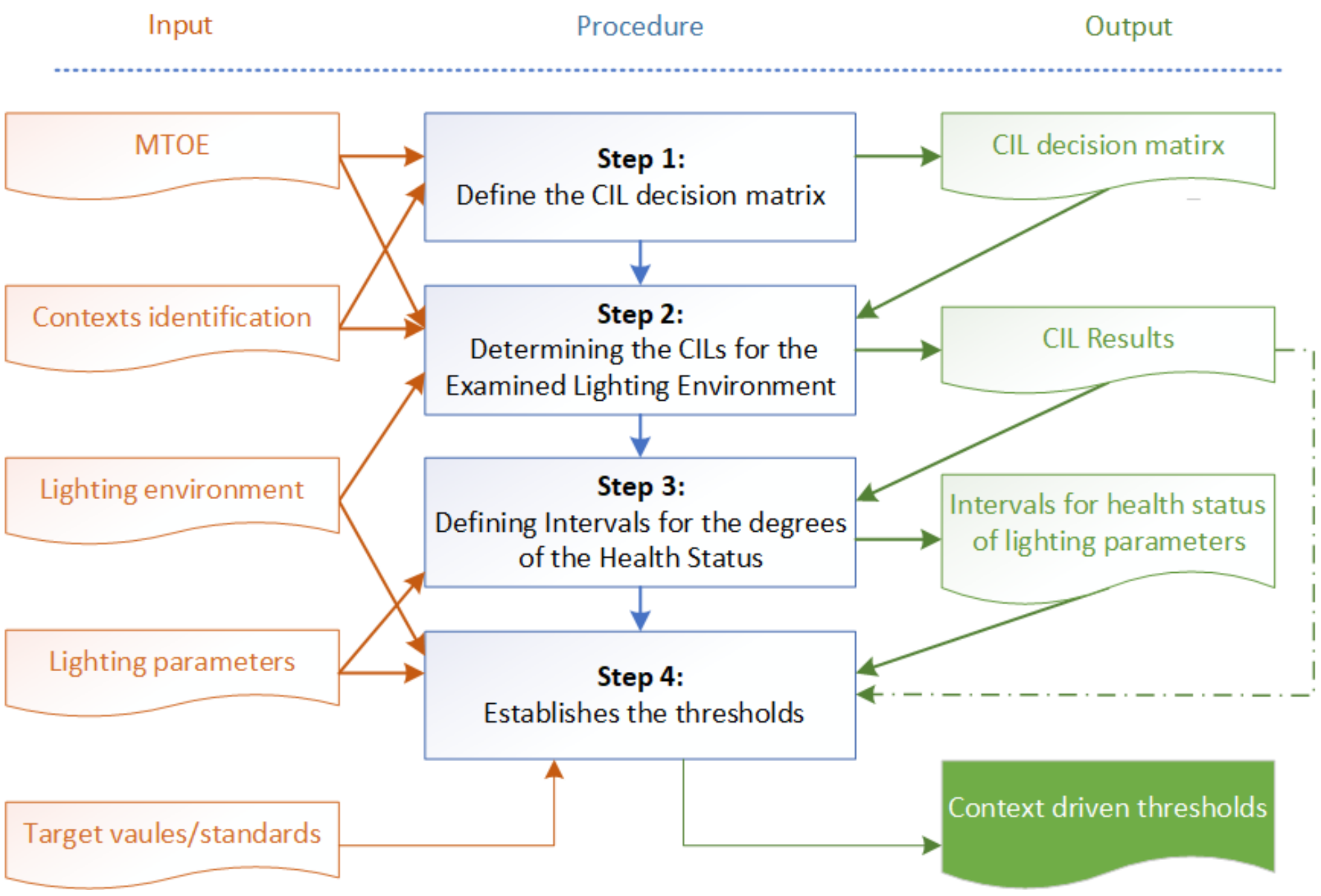

Fig.1 Input, output, and main steps in the procedure of setting context driven thresholds

**Step 1: Define the CIL Matrix** - The CIL matrix is formulated based on Mean Time of Exposure (MTOE) and the identified contexts pertinent to the study. For this research, the context in focus is the age of the users. According to Table 3.1, users under 12 years of age are categorized into Group A, those over 65 into Group C, and all other users into Group B. Following Assumption 1 from section 3.1, an MTOE of less than 5 minutes is classified as T1, more than 15 minutes as T3, and durations in between as T2. Within this framework, CIL-I denotes a higher priority level than CIL-II. Specifically, CIL-I is assigned to Group A - T3, indicating a high Critical Integrative Level for lighting environments predominantly used by users younger than 12 for periods longer than 15 minutes, such as a children's reading area.

Table 3.1 Context driven CIL matrix: an example.

| MTOE/Ages | Younger than 12 (Group A) | Between (12, 65) (Group B) | Older than 65 (Group C) |
|---|---|---|---|
| Less than 5 (T1) | CIL-II | CIL-II | CIL-II |
| Between [5 – 15] (T2) | CIL-I | CIL-II | CIL-I |
| Longer than 15 (T3) | CIL-I | CIL-I | CIL-I |

**Step 2: Determining the CILs for the Examined Lighting Environment** – This step involves identifying the Critical Integrative Levels (CILs) across the entire lighting environment under study, which, in this case, is the public library. As illustrated in Table 3.2, this process entails analysing the specific sub-environments of lighting assets within the library and their primary users, identifying the age groups of these users and their Mean Time of Exposure (MTOE). Utilizing the outcomes from the previous step, the appropriate CILs for each sub-environment are then established based on these factors.

**Step 3: Defining Intervals for the degrees of the Health Status Using the CIL Matrix** - The health status of each lighting parameter (as discussed in Section 2.1) can be classified into varying degrees based on the CIL matrix. These degrees, corresponding to different CILs, determine the interval or range for each health status. An illustrative example is provided in Table 3.3, where, for instance, the health status (HS) of illuminance is categorized into five degrees, from HS1 to HS5, with HS5 being the optimal status. In the table, a 0% indication means that the actual value meets the requirements of standard limits (S).

Target (T) values determined by the designer, which are set to achieve higher levels of lighting performance. It is important to recognize that exceeding necessary lighting values can lead to resource inefficiency, therefore the optimal achievable status is HS4, not HS5, to prevent wasteful over-illumination. Detailed intervals for the health status of lighting parameters are discussed in Chapter 4. For parameters where a higher level indicates superior performance, such as HS5 being better than HS4, HS5 is highlighted in dark blue.

Table 3.2 Identified CILs with specified lighting environments: some examples.

| Objectives_application_main users | Age group | MTOE | CIL |
|---|---|---|---|
| Bookshelf_Display_Adult | Group A, B, C | T1 | CIL-II |
| Bookshelf_Reserve_Adult | Group B, C | T2 | CIL-I |
| Table_Book borrow_digital_Adult | Group B | T1 | CIL-II |
| Table_Book return_ Adult | Group A, B, C | T1 | CIL-II |
| Table Counter_multifunction_Staff | Group B | T3 | CIL-I |
| Sofa_ Reading/learning_Adult | Group C | T3 | CIL-I |
| Sofa_ Reading/learning_Children | Group A | T2 | CIL-I |
| Table_Digital inquiry_Adult | Group B | T1 | CIL-II |
| Table_Reading/learning_multifunction_Adult | Group B | T3 | CIL-I |
| Table_Reading/learning_digital_Adult | Group B | T3 | CIL-I |
| Table_ Hobby_Children | Group A | T2 | CIL-I |
| Table_ Workshop_Adult | Group C | T3 | CIL-I |

Table 3.3 Intervals for health status of lighting parameters: an example of illuminance $\bar{E}_m$.

| Visualization by colours | Health status (HS) degree | CIL-I | CIL-II |
|---|---|---|---|
| | HS1 | [0, 95%*S) | [0, 90%*S) |
| | HS2 | [95%*S, S) | [90%*S, S) |
| | HS3 | [S, T) | [S, T) |
| | HS4 | [T, 150%*T] | [T, 120%T] |
| | HS5 | (150%*T, $+\infty$) | (120%T, $+\infty$) |

**Step 4 Establishes the thresholds based on contact driven CIL matrix for the examined lighting environment** – Based on the health status intervals set up in Step 3 and the target values for key lighting parameters in specified lighting environment, establishes the thresholds for each status. Take illuminance as example (see Table 3.3), for environments classified under CIL-I, the health status range is set to HS5, with the parameter's target value falling within a range of [T, 150%*T). Conversely, for CIL-II environments, the range is narrowed to [T, 120%*T). Target (T) values are typically established during the design phase by the lighting designers; however, in the absence of specific design data, standard values are used as benchmarks. This differentiation underscores the higher priority of CIL-I over CIL-II, reflected through broader health status intervals. Table 3.4 presents some examples of setting thresholds of $\bar{E}_m$ in the public library based on the above CIL matrix in Table 3.3, in which we assume the Target Value (T) is about 20% higher than the Standard value (S).

Table 3.4 Thresholds of $\bar{E}_m$: some examples with S=300, T=400

| Main Objectives_Users | CIL | Standards/ Target | Health Status | | | | |
|---|---|---|---|---|---|---|---|
| | | | HS1 | HS2 | HS3 | **HS5** | HS4 |
| Bookshelf_Display_Adult | II | 300/400 | (0, 270) | [270, 300) | [300, 360) | [360, 480] | (480, $+\infty$) |
| Bookshelf_Reserve_Adult | I | 300/400 | (0, 285) | [285, 300) | [300, 360) | [360, 600] | (600, $+\infty$) |

## 4. A Guide on Context driven Thresholds for Key Lighting Parameters

Following the approaches proposed in chapter 3, Chapter 4 provides the thresholds established in the project for recommended key lighting parameters for monitoring and diagnostics. The recommended parameters are selected from the identified key parameters in Chapter 2 (Table.2.1), which provides a guide on setting context driven thresholds considering the needs on human centric and integrative lighting, requirement of standards, as well as target value from the lighting designers. These parameters (listed in Table.4.1) are recommended for lighting performance measurement in public libraries for the purpose of monitoring and diagnostics. Some of the applications will be illustrated with case studies in Chapter 5.

Table. 4.1 Recommended lighting parameters for monitoring and diagnostics

| Category | Sub-category | Symbol |
|---|---|---|
| Illuminance | Maintained average illuminance | $\bar{E}_m$ |
| | Uniformity ratio of illuminance | $U_0$ |
| Glare | Unified Glare Rating | $UGR_L$ |
| Colour Rendering | Colour rendering Index $R_a$ | $R_a$ |
| | Colour rendering Index $R_9$ | $R_9$ |
| | TM-30 fidelity index | $R_f$ |
| | TM-30 gamut area | $R_g$ |
| | MACADAM | SDCM |
| Colour Temperaturer | Correlated Colour Temperaturer | CCT |
| | Delta Ultraviolet | DUV |
| Flicker | Stroboscopic Visibility Measure | SVM |
| Circadian Rhythm | Equivalent Melanopic Lux | EML |

### 4.1 A guide on thresholds of illuminance: $\bar{E}_m$ and $U_0$

Illumination parameters should be in proximity of Target Value (T) to ensure a balanced lighting design throughout the library. Higher levels indicated as HS4 in $\bar{E}_m$will lessen energy efficiency. Based on Standard SS-EN 12464-1 [14], Table 3.2 and Table 3.3, a guide on context driven threshold of $\bar{E}_m$ can be found in Table 4.1.1., in which we assume the target (T) value, where T ($\bar{E}_m$)= S ($\bar{E}_m$)+100, as shown in the Tabel 4.1.2. For $U_0$, we assume that T ($U_0$)=S ($U_0$)+0.1. A $U_0$ of 1, which eliminates all shadows, could create a very clinical and monotonous atmosphere if applied throughout an entire library. Currently, we are using higher ratios only in specific areas such as sofas or tables to enhance those spaces. It's important to clarify that we are not advocating for a uniformity ratio of 1 across the entire library, as this might not provide the best aesthetic or functional outcome.

### 4.2 A guide on thresholds of glare: $UGR_L$

Glare needs to be avoided to ensure a good working environment. A glare rating lower than standard is desired. Based on Standard SS-EN 12464-1 [14], the variables are between 10 (best) -30 (worst). A guide on context driven threshold of $UGR_L$ can be found in Table 4.2.1.

### 4.3 A guide on thresholds of colour Rendering: $R_a, R_9, R_f, R_g$, SDCM

The higher colour rendering, the better environment in the library. Colour rendering above target remains positive. Based on Standard SS-EN 12464-1 [14] and/or target, the variables $R_a, R_9, R_f$, are between 1-100, , $R_g$ is between 1-150. A guide on context driven threshold of $R_a, R_9, R_f, R_g$, SDCM, can be found in Table 4.3.1-Tabele 4.3.5.

### 4.4 A guide on thresholds of colour temperature: CCT and DUV

The colour appearance of a light source refers to the apparent colour (chromaticity) of the light emitted. It is quantified by its correlated colour temperature (CCT). The choice of colour appearance of the light is a matter of psychology, aesthetics and what is preferred. Lighting sources need to stay in proximity to target to ensure a pleasant environment. The CCT is regulated by the lighting designer. While standard such as SS-EN 12464-1 [14] has only recommendations for special conditions (not for libraries). In this study, we make suggestions including:

- For General Reading Areas, the average CCT could be 3500 K. The reason is that a neutral to cool white colour temperature helps in maintaining alertness and reduces eye strain, which is beneficial for reading and writing.
- For Children's Sections, the average CCT could be 3300K. The reason is that a warmer colour temperature creates a welcoming and stimulating environment for children, which can be more comfortable and less harsh for young eyes.
- For Computer and Learning Table Areas: the average CCT could be 4000 K. The reason is that cooler white light is preferable where tasks require concentration and detail (such as computer use and research), as it promotes alertness and efficiency.
- For Relaxation Zones (Sofa, etc.): the average CCT could be 3000K. The reason is that warm white lighting can make those areas more inviting and relaxing, making it easier for people to unwind with a book.

Based on the above suggestions, a guide on context driven threshold of CCT can be found in Table 4.4.1.

The DUV value that is considered acceptable for human perception and application requirements can vary depending on the context and the specific standards applied. In general, for most indoor lighting applications where colour quality is important, such as in residential or office settings, a DUV value close to zero is preferred, typically within the range of -0.006 to +0.006 [15]. In this study, a guide on context driven threshold of CCT can be found in Table 4.4.2. This range ensures that the light colour is perceived as neutral white, without noticeable green or magenta tints that can be objectionable or cause discomfort over time.

### 4.5 A guide on thresholds of flicker parameter: SVM

SVM, the metric used to quantify the risk of stroboscopic effects that can be perceived by humans in a flickering light source. This is particularly relevant with the widespread use of LED lighting, which, if not properly controlled, can exhibit pronounced flicker. SVM values typically range from 0 to 1, with 0 indicating no visible stroboscopic effect and 1 indicating a highly noticeable effect. For library environments, where focused activities such as reading and computer work take place, it is important to minimize any potential for distraction or discomfort. Based on the Eco-design directive for regulation of maximum value [11], a guide on context driven threshold of CCT can be found in Table 4.4.2.

### 4.7 Equivalent Melanopic Lux (EML)

The higher EML, the healthier contribution to the visual environment in the library. In certain settings, some positive compensations will differ from the dynamics in natural lighting environment. Based on WELL Standard [16], American National Standards RP-1-12 [17], British Standards Institution PD CEN/TR 16791:2017 [18], DIN/TS 5031-100:2021-11 [19], CIE S 026:2018 [20], A guide on context driven threshold of EML can be found in Table 4.5.1. Noted, this study takes recommended thresholds from WELL V2 in which EML is recommended to be>150 EML, while DIN standard generally recommends >250 EML.

Table 4.1.1 Intervals and Context driven thresholds of $\bar{E}_m$ compared to Standard (S) and Target (T) value.

| Visualization by colours | Health status degree | CIL-I | CIL-II |
|---|---|---|---|
| 🟥 | HS1 | [0, 95%*S) | [0, 90%*S) |
| 🟧 | HS2 | [95%*S, S) | [90%*S, S) |
| 🟨 | HS3 | [S, T) | [S, T) |
| 🟩 | HS4 | [T, 150%*T] | [T, 120%T] |
| 🟩 | HS5 | (150%*T, $+\infty$) | (120%T, $+\infty$) |

| Objectives/thresholds | CIL | S/T | HS1 | HS2 | HS3 | HS4 | HS5 |
|---|---|---|---|---|---|---|---|
| Bookshelf_Display_Adult/Children | II | 300/400 | (0, 270) | [270, 300) | [300, 360) | [360, 480] | (480, $+\infty$) |
| Bookshelf_Reserve_Adult/Children | I | 300/400 | (0, 285) | [285, 300) | [300, 360) | [360, 600] | (600, $+\infty$) |
| Table_Book borrow_digital_Adult | II | 500/600 | (0, 450) | [450, 500) | [500, 600) | [600, 720] | (720, $+\infty$) |
| Table_Book return_ Adult | II | 500/600 | (0, 450) | [450, 500) | [500, 600) | [600, 720] | (720, $+\infty$) |
| Table Counter_multifunction_Staff | I | 750/850 | (0, 713) | [713, 750) | [750, 850) | [850, 1275] | (1275, $+\infty$) |
| Sofa_ Reading/learning_Adult | I | 750/850 | (0, 713) | [713, 750) | [750, 850) | [850, 1275] | (1275, $+\infty$) |
| Sofa_ Reading/learning_Children | I | 500/600 | (0, 475) | [475, 500) | [500, 600) | [600, 900] | (900, $+\infty$) |
| Table_Digital inquiry_Adult | II | 500/600 | (0, 450) | [450, 500) | [500, 600) | [600, 720] | (720, $+\infty$) |
| Table_Reading/learning_multifunction_Adult | I | 750/850 | (0, 713) | [713, 750) | [750, 850) | [850, 1275] | (1275, $+\infty$) |
| Table_Reading/learning_digital_Adult | I | 500/600 | (0, 475) | [475, 500) | [500, 600) | [600, 900] | (900, $+\infty$) |
| Table_ Hobby_Children | I | 500/600 | (0, 475) | [475, 500) | [500, 600) | [600, 900] | (900, $+\infty$) |
| Table_ Workshop_Adult | I | 750/850 | (0, 713) | [713, 750) | [750, 850) | [850, 1275] | (1275, $+\infty$) |

Table 4.1.2 Intervals and Context driven thresholds of $U_0$ compared to Standard (S) and Target (T) value.

| Visualization by colours | Health status degree | CIL-I | CIL-II |
|---|---|---|---|
| 🟥 | HS1 | [0, 95%*S) | [0, 90%*S) |
| 🟧 | HS2 | [95%*S, S) | [90%*S, S) |
| 🟨 | HS3 | [S, T) | [S, T) |
| 🟩 | HS4 | [T, 120%*T] | [T, 110%*T] |
| 🟩 | HS5 | (120%*T, 1) | (110%*T, 1) |

| Objectives/thresholds | CIL | S/T | HS1 | HS2 | HS3 | HS4 | HS5 |
|---|---|---|---|---|---|---|---|
| Bookshelf_Display_Adult | II | 0.4/0.5 | (0, 0.36) | [0.36, 0.4) | [0.4, 0.5) | [0.5, 0.55] | (0.55, 1) |
| Bookshelf_Reserve_Adult | I | 0.4/0.5 | (0, 0.38) | [0.38, 0.4) | [0.4, 0.5) | [0.5, 0.6] | (0.6, 1) |
| Table_Book borrow_digital_Adult | II | 0.6/0.7 | (0, 0.54) | [0.54, 0.6) | [0.6, 0.7) | [0.7, 0.77] | (0.77, 1) |
| Table_Book return_ Adult | II | 0.6/0.7 | (0, 0.54) | [0.54, 0.6) | [0.6, 0.7) | [0.7, 0.77] | (0.77, 1) |
| Table Counter_multifunction_Staff | I | 0.6/0.7 | (0, 0.547) | [0.57, 0.6) | [0.6, 0.7) | [0.7, 0.84] | (0.84, 1) |
| Sofa_ Reading/learning_Adult | I | 0.6/0.7 | (0, 0.547) | [0.57, 0.6) | [0.6, 0.7) | [0.7, 0.84] | (0.84, 1) |
| Sofa_ Reading/learning_Children | I | 0.6/0.7 | (0, 0.547) | [0.57, 0.6) | [0.6, 0.7) | [0.7, 0.84] | (0.84, 1) |
| Table_Digital inquiry_Adult | II | 0.6/0.7 | (0, 0.54) | [0.54, 0.6) | [0.6, 0.7) | [0.7, 0.77] | (0.77, 1) |
| Table_Reading/learning_multifunction_Adult | I | 0.6/0.7 | (0, 0.547) | [0.57, 0.6) | [0.6, 0.7) | [0.7, 0.84] | (0.84, 1) |
| Table_Reading/learning_digital_Adult | I | 0.6/0.7 | (0, 0.547) | [0.57, 0.6) | [0.6, 0.7) | [0.7, 0.84] | (0.84, 1) |
| Table_ Hobby_Children | I | 0.6/0.7 | (0, 0.547) | [0.57, 0.6) | [0.6, 0.7) | [0.7, 0.84] | (0.84, 1) |
| Table_ Workshop_Adult | I | 0.6/0.7 | (0, 0.547) | [0.57, 0.6) | [0.6, 0.7) | [0.7, 0.84] | (0.84, 1) |

Table 4.2.1 Intervals and Context driven thresholds of $UGR_L$ compared to Standard (S) value.

| Visualization by colours | Health status degree | CIL-I | CIL-II |
|---|---|---|---|
| | HS1 | [S+1.5, +∞) | [S+3, +∞) |
| | HS2 | [S+1.2, S+1.5) | [S+1.5, S+3) |
| | HS3 | [S, S+1.3) | [S, S+1.5) |
| | HS4 | [S-3, S] | [S-2, S] |
| | HS5 | (0, S-3) | (0, S-2) |

| Objectives/thresholds | CIL | S | HS1 | HS2 | HS3 | HS4 | HS5 |
|---|---|---|---|---|---|---|---|
| Bookshelf_Display_Adult | II | 19 | (22, +∞) | [20.5, 22) | [19, 20.5) | [17, 19] | (0, 17) |
| Bookshelf_Reserve_Adult | I | 19 | (20.5, +∞) | [20.2, 20.5) | [19, 20.2) | [16, 19] | (0, 16) |
| Table_Book borrow_digital_Adult | II | 22 | (25, +∞) | [23.5, 25) | [22, 23.5) | [20, 22] | (0, 20) |
| Table_Book return_ Adult | II | 22 | (25, +∞) | [23.5, 25) | [22, 23.5) | [20, 22] | (0, 20) |
| Table Counter_multifunction_Staff | I | 19 | (20.5, +∞) | [20.2, 20.5) | [19, 20.2) | [16, 19] | (0, 16) |
| Sofa_ Reading/learning_Adult | I | 19 | (20.5, +∞) | [20.2, 20.5) | [19, 20.2) | [16, 19] | (0, 16) |
| Sofa_ Reading/learning_Children | I | 19 | (20.5, +∞) | [20.2, 20.5) | [19, 20.2) | [16, 19] | (0, 16) |
| Table_Digital inquiry_Adult | II | 19 | (22, +∞) | [20.5, 22) | [19, 20.5) | [17, 19] | (0, 17) |
| Table_Reading/learning_multifunction_Adult | I | 19 | (20.5, +∞) | [20.2, 20.5) | [19, 20.2) | [16, 19] | (0, 16) |
| Table_Reading/learning_digital_Adult | I | 19 | (20.5, +∞) | [20.2, 20.5) | [19, 20.2) | [16, 19] | (0, 16) |
| Table_ Hobby_Children | I | 19 | (20.5, +∞) | [20.2, 20.5) | [19, 20.2) | [16, 19] | (0, 16) |
| Table_ Workshop_Adult | I | 19 | (20.5, +∞) | [20.2, 20.5) | [19, 20.2) | [16, 19] | (0, 16) |

Table 4.3.1 Intervals for health status of CRI $R_a$ compared to target and standard SS-EN 12464-1 (T=S+5)

| Visualization by colours | Health status degree | CIL-I | CIL-II |
|---|---|---|---|
| | HS1 | [0, 95%*S) | [0, 90%*S) |
| | HS2 | [95%*S, S) | [90%*S, S) |
| | HS3 | [S, T) | [S, T) |
| | HS4 | [T, 115%*T] | [T, 110%*T] |
| ] | HS5 | (115%*T, 100) | (110%*T, 100) |

| Objectives/thresholds | CIL | S/T | HS1 | HS2 | HS3 | HS4 | HS5 |
|---|---|---|---|---|---|---|---|
| Bookshelf_Display_Adult | II | 80/85 | (0, 72) | [72, 80) | [80, 85) | [85, 93.5] | (93.5, 100) |
| Bookshelf_Reserve_Adult | I | 80/85 | (0, 76) | [76, 80) | [80, 85) | [85, 97.75] | (97.75, 100) |
| Table_Book borrow_digital_Adult | II | 80/85 | (0, 72) | [72, 80) | [80, 85) | [85, 93.5] | (93.5, 100) |
| Table_Book return_ Adult | II | 80/85 | (0, 72) | [72, 80) | [80, 85) | [85, 93.5] | (93.5, 100) |
| Table Counter_multifunction_Staff | I | 80/85 | (0, 76) | [76, 80) | [80, 85) | [85, 97.75] | (97.75, 100) |
| Sofa_ Reading/learning_Adult | I | 80/85 | (0, 76) | [76, 80) | [80, 85) | [85, 97.75] | (97.75, 100) |
| Sofa_ Reading/learning_Children | I | 80/85 | (0, 76) | [76, 80) | [80, 85) | [85, 97.75] | (97.75, 100) |
| Table_Digital inquiry_Adult | II | 80/85 | (0, 72) | [72, 80) | [80, 85) | [85, 93.5] | (93.5, 100) |
| Table_Reading/learning_multifunction_Adult | I | 80/85 | (0, 76) | [76, 80) | [80, 85) | [85, 97.75] | (97.75, 100) |
| Table_Reading/learning_digital_Adult | I | 80/85 | (0, 76) | [76, 80) | [80, 85) | [85, 97.75] | (97.75, 100) |
| Table_ Hobby_Children | I | 80/85 | (0, 76) | [76, 80) | [80, 85) | [85, 97.75] | (97.75, 100) |
| Table_ Workshop_Adult | I | 80/85 | (0, 76) | [76, 80) | [80, 85) | [85, 97.75] | (97.75, 100) |

Table 4.3.2 Intervals for health status of CRI $R_9$ compared to target (T=60)

| Visualization by colours | Health status degree | CIL-I | CIL-II |
|---|---|---|---|
| | HS1 | [0, 50%*T) | [0, 40%*T) |
| | HS2 | [50%*T, 65%*T) | [40%*T, 55%*T) |
| | HS3 | [65%*T, 80%*T) | [55%*T, 70%*T) |
| | HS4 | [80%*T, T) | [70%*T, T) |
| | HS5 | [T, 100) | [T, 100) |

| Objectives/thresholds | CIL | T | HS1 | HS2 | HS3 | HS4 | HS5 |
|---|---|---|---|---|---|---|---|
| Bookshelf_Display_Adult | II | 60 | [0, 24) | [24, 33) | [33, 42) | [42, 60) | [60, 100) |
| Bookshelf_Reserve_Adult | I | 60 | [0, 30) | [30, 39) | [39, 48) | [48, 60) | [60, 100) |
| Table_Book borrow_digital_Adult | II | 60 | [0, 24) | [24, 33) | [33, 42) | [42, 60) | [60, 100) |
| Table_Book return_ Adult | II | 60 | [0, 24) | [24, 33) | [33, 42) | [42, 60) | [60, 100) |
| Table Counter_multifunction_Staff | I | 60 | [0, 30) | [30, 39) | [39, 48) | [48, 60) | [60, 100) |
| Sofa_ Reading/learning_Adult | I | 60 | [0, 30) | [30, 39) | [39, 48) | [48, 60) | [60, 100) |
| Sofa_ Reading/learning_Children | I | 60 | [0, 30) | [30, 39) | [39, 48) | [48, 60) | [60, 100) |
| Table_Digital inquiry_Adult | II | 60 | [0, 24) | [24, 33) | [33, 42) | [42, 60) | [60, 100) |
| Table_Reading/learning_multifunction_Adult | I | 60 | [0, 30) | [30, 39) | [39, 48) | [48, 60) | [60, 100) |
| Table_Reading/learning_digital_Adult | I | 60 | [0, 30) | [30, 39) | [39, 48) | [48, 60) | [60, 100) |
| Table_ Hobby_Children | I | 60 | [0, 30) | [30, 39) | [39, 48) | [48, 60) | [60, 100) |
| Table_ Workshop_Adult | I | 60 | [0, 30) | [30, 39) | [39, 48) | [48, 60) | [60, 100) |

Table 4.3.3 Intervals for health status of CRI $R_f$ compared to target and standard SS-EN 12464-1 (T=S+5)

| Visualization by colours | Health status degree | CIL-I | CIL-II |
|---|---|---|---|
| | HS1 | [0, 95%*S) | [0, 90%*S) |
| | HS2 | [95%*S, S) | [90%*S, S) |
| | HS3 | [S, T) | [S, T) |
| | HS4 | [T, 115%*T] | [T, 110%*T] |
| | HS5 | (115%*T, 100) | (110%*T, 100) |

| Objectives/thresholds | CIL | S/T | HS1 | HS2 | HS3 | HS4 | HS5 |
|---|---|---|---|---|---|---|---|
| Bookshelf_Display_Adult | II | 80/85 | (0, 72) | [72, 80) | [80, 85) | [85, 93.5] | (93.5, 100) |
| Bookshelf_Reserve_Adult | I | 80/85 | (0, 76) | [76, 80) | [80, 85) | [85, 97.75] | (97.75, 100) |
| Table_Book borrow_digital_Adult | II | 80/85 | (0, 72) | [72, 80) | [80, 85) | [85, 93.5] | (93.5, 100) |
| Table_Book return_ Adult | II | 80/85 | (0, 72) | [72, 80) | [80, 85) | [85, 93.5] | (93.5, 100) |
| Table Counter_multifunction_Staff | I | 80/85 | (0, 76) | [76, 80) | [80, 85) | [85, 97.75] | (97.75, 100) |
| Sofa_ Reading/learning_Adult | I | 80/85 | (0, 76) | [76, 80) | [80, 85) | [85, 97.75] | (97.75, 100) |
| Sofa_ Reading/learning_Children | I | 80/85 | (0, 76) | [76, 80) | [80, 85) | [85, 97.75] | (97.75, 100) |
| Table_Digital inquiry_Adult | II | 80/85 | (0, 72) | [72, 80) | [80, 85) | [85, 93.5] | (93.5, 100) |
| Table_Reading/learning_multifunction_Adult | I | 80/85 | (0, 76) | [76, 80) | [80, 85) | [85, 97.75] | (97.75, 100) |
| Table_Reading/learning_digital_Adult | I | 80/85 | (0, 76) | [76, 80) | [80, 85) | [85, 97.75] | (97.75, 100) |
| Table_ Hobby_Children | I | 80/85 | (0, 76) | [76, 80) | [80, 85) | [85, 97.75] | (97.75, 100) |
| Table_ Workshop_Adult | I | 80/85 | (0, 76) | [76, 80) | [80, 85) | [85, 97.75] | (97.75, 100) |

Table 4.3.4 Intervals for health status of CRI $R_g$ compared to target (T=95)

| Visualization by colours | Health status degree | CIL-I | CIL-II |
|---|---|---|---|
| | HS1 | [1, 85]; [110, 150) | [1, 80]; [115, 150) |
| | HS2 | (85, 90]; [105, 110] | (80, 85]; [110, 115] |
| | HS3 | (90, 95]; [103, 105] | (85, 95]; [103, 110] |
| | HS4 | (95, 99]; [101, 103] | (95, 99]; [101, 103] |
| | HS5 | (99, 101) | (99, 101) |

| Objectives/thresholds | CIL | T | HS1 | HS2 | HS3 | HS4 | HS5 |
|---|---|---|---|---|---|---|---|
| Bookshelf_Display_Adult | II | 0 | [1, 80]; [115, 150) | (80, 85]; [110, 115] | (85, 95]; [103, 110] | (95, 99]; [101, 103] | (99, 101) |
| Bookshelf_Reserve_Adult | I | 0 | [1, 85]; [110, 150) | (85, 90]; [105, 110] | (90, 95]; [103, 105] | | |
| Table_Book borrow_digital_Adult | II | 0 | [1, 80]; [115, 150) | (80, 85]; [110, 115] | (85, 95]; [103, 110] | | |
| Table_Book return_ Adult | II | 0 | [1, 80]; [115, 150) | (80, 85]; [110, 115] | (85, 95]; [103, 110] | | |
| Table Counter_multifunction_Staff | I | 0 | [1, 85]; [110, 150) | (85, 90]; [105, 110] | (90, 95]; [103, 105] | | |
| Sofa_ Reading/learning_Adult | I | 0 | [1, 85]; [110, 150) | (85, 90]; [105, 110] | (90, 95]; [103, 105] | | |
| Sofa_ Reading/learning_Children | I | 0 | [1, 85]; [110, 150) | (85, 90]; [105, 110] | (90, 95]; [103, 105] | | |
| Table_Digital inquiry_Adult | II | 0 | [1, 80]; [115, 150) | (80, 85]; [110, 115] | (85, 95]; [103, 110] | | |
| Table_Reading/learning_multifunction_Adult | I | 0 | [1, 85]; [110, 150) | (85, 90]; [105, 110] | (90, 95]; [103, 105] | | |
| Table_Reading/learning_digital_Adult | I | 0 | [1, 85]; [110, 150) | (85, 90]; [105, 110] | (90, 95]; [103, 105] | | |
| Table_ Hobby_Children | I | 0 | [1, 85]; [110, 150) | (85, 90]; [105, 110] | (90, 95]; [103, 105] | | |
| Table_ Workshop_Adult | I | 0 | [1, 85]; [110, 150) | (85, 90]; [105, 110] | (90, 95]; [103, 105] | | |

Table 4.3.5 Intervals for health status of SDCM compared to target (T=3)

| Visualization by colours | Health status degree | CIL-I | CIL-II |
|---|---|---|---|
| | HS1 | [5, +∞) | [6, +∞) |
| | HS2 | [4, 5) | [5, 6) |
| | HS3 | [3, 4) | [3, 5) |
| | HS4 | [2, 3) | [2, 3) |
| ] | HS5 | (1, 2) | (1, 2) |

| Objectives/thresholds | CIL | T | HS1 | HS2 | HS3 | HS4 | HS5 |
|---|---|---|---|---|---|---|---|
| Bookshelf_Display_Adult | II | 3 | [6, +∞) | [5, 6) | [3, 5) | [2, 3] | (1, 2) |
| Bookshelf_Reserve_Adult | I | 3 | [5, +∞) | [4, 5) | [3, 4) | [2, 3] | (1, 2) |
| Table_Book borrow_digital_Adult | II | 3 | [6, +∞) | [5, 6) | [3, 5) | [2, 3] | (1, 2) |
| Table_Book return_ Adult | II | 3 | [6, +∞) | [5, 6) | [3, 5) | [2, 3] | (1, 2) |
| Table Counter_multifunction_Staff | I | 3 | [5, +∞) | [4, 5) | [3, 4) | [2, 3] | (1, 2) |
| Sofa_ Reading/learning_Adult | I | 3 | [5, +∞) | [4, 5) | [3, 4) | [2, 3] | (1, 2) |
| Sofa_ Reading/learning_Children | I | 3 | [5, +∞) | [4, 5) | [3, 4) | [2, 3] | (1, 2) |
| Table_Digital inquiry_Adult | II | 3 | [6, +∞) | [5, 6) | [3, 5) | [2, 3] | (1, 2) |
| Table_Reading/learning_multifunction_Adult | I | 3 | [5, +∞) | [4, 5) | [3, 4) | [2, 3] | (1, 2) |
| Table_Reading/learning_digital_Adult | I | 3 | [5, +∞) | [4, 5) | [3, 4) | [2, 3] | (1, 2) |
| Table_ Hobby_Children | I | 3 | [5, +∞) | [4, 5) | [3, 4) | [2, 3] | (1, 2) |
| Table_ Workshop_Adult | I | 3 | [5, +∞) | [4, 5) | [3, 4) | [2, 3] | (1, 2) |

Table 4.4.1 Intervals and Context driven thresholds of CCT compared to Target (T) value.

| Visualization by colours | Health status degree | CIL-I | CIL-II |
|---|---|---|---|
| | HS1 | [0, 60%*T) | [0, 50%*T) |
| | HS2 | [60%*T, 90%*T) | [50%*T, 80%*T) |
| | HS3 | [90%*T, 95%T) | [80%*T, 95%*T) |
| | HS4 | [95%*T, 105%*T) | [95%*T, 110%*T) |
| | HS5 | [105%*T, +∞) | [110%*T, +∞) |

| Objectives/thresholds | CIL | T | HS1 | HS2 | HS3 | HS4 | HS5 |
|---|---|---|---|---|---|---|---|
| Bookshelf_Display_Adult | II | 3500 | (0, 1750) | [1750, 2800) | [2800, 3325) | [3325, 3850) | [3850, +∞) |
| Bookshelf_Reserve_Adult | I | 3500 | (0, 2100) | [2100, 3150) | [3150, 3325) | [3325, 3675) | [3675, +∞) |
| Table_Book borrow_digital_Adult | II | 3500 | (0, 1750) | [1750, 2800) | [2800, 3325) | [3325, 3850) | [3850, +∞) |
| Table_Book return_ Adult | II | 3500 | (0, 1750) | [1750, 2800) | [2800, 3325) | [3325, 3850) | [3850, +∞) |
| Table Counter_multifunction_Staff | I | 4000 | (0, 2400) | [2400, 3600) | [3600, 3800) | [3800, 4200) | [4200, +∞) |
| Sofa_ Reading/learning_Adult | I | 3000 | (0, 1800) | [1800, 2700) | [2700, 2850) | [2850, 3150) | [3150, +∞) |
| Sofa_ Reading/learning_Children | I | 3000 | (0, 1800) | [1800, 2700) | [2700, 2850) | [2850, 3150) | [3150, +∞) |
| Table_Digital inquiry_Adult | II | 4000 | (0, 2000) | [2000, 3200) | [3200, 3800) | [3800, 4400) | [5200, +∞) |
| Table_Reading/learning_multifunction_Adult | I | 4000 | (0, 2400) | [2400, 3600) | [3600, 3800) | [3800, 4200) | [4200, +∞) |
| Table_Reading/learning_digital_Adult | I | 4000 | (0, 2400) | [2400, 3600) | [3600, 3800) | [3800, 4200) | [4200, +∞) |
| Table_ Hobby_Children | I | 3300 | (0, 1980) | [1980, 2970) | [2970, 3135) | [3135, 3465) | (3465, +∞) |
| Table_ Workshop_Adult | I | 4000 | (0, 2400) | [2400, 3600) | [3600, 3800) | [3800, 4200) | [4200, +∞) |

Table 4.4.2 Intervals and Context driven thresholds of DUV compared to Target (T) value.

| Visualization by colours | Health status degree | CIL-I | CIL-II |
|---|---|---|---|
| | HS1 | (-∞, -0.005]; [0.005, +∞) | (-∞, -0.006]; [0.006, +∞) |
| | HS2 | (-0.005, -0.004]; [0.004, 0.005] | (-0.006, -0.005]; [0.005, 0.006] |
| | HS3 | (-0.004, -0.003]; [0.003, 0.004] | (-0.005, -0.003]; [0.003, 0.005] |
| | HS4 | (-0.003, -0.001]; [0.001, 0.003] | (-0.003, -0.001]; [0.001, 0.003] |
| | HS5 | (-0.001, 0.001) | (-0.001, 0.001) |

| Objectives/thresholds | CIL | T | HS1 | HS2 | HS3 | HS4 | HS5 |
|---|---|---|---|---|---|---|---|
| Bookshelf_Display_Adult | II | 0 | (-∞, -0.006]; [0.006, +∞) | (-0.006, -0.005]; [0.005, 0.006] | (-0.005, -0.003]; [0.003, 0.005] | (-0.003, -0.001]; [0.001, 0.003] | (-0.001, 0.001) |
| Bookshelf_Reserve_Adult | I | 0 | (-∞, -0.005]; [0.005, +∞) | (-0.005, -0.004]; [0.004, 0.005] | (-0.004, -0.003]; [0.003, 0.004] | | |
| Table_Book borrow_digital_Adult | II | 0 | (-∞, -0.006]; [0.006, +∞) | (-0.006, -0.005]; [0.005, 0.006] | (-0.005, -0.003]; [0.003, 0.005] | | |
| Table_Book return_ Adult | II | 0 | (-∞, -0.006]; [0.006, +∞) | (-0.006, -0.005]; [0.005, 0.006] | (-0.005, -0.003]; [0.003, 0.005] | | |
| Table Counter_multifunction_Staff | I | 0 | (-∞, -0.005]; [0.005, +∞) | (-0.005, -0.004]; [0.004, 0.005] | (-0.004, -0.003]; [0.003, 0.004] | | |
| Sofa_ Reading/learning_Adult | I | 0 | (-∞, -0.005]; [0.005, +∞) | (-0.005, -0.004]; [0.004, 0.005] | (-0.004, -0.003]; [0.003, 0.004] | | |
| Sofa_ Reading/learning_Children | I | 0 | (-∞, -0.005]; [0.005, +∞) | (-0.006, -0.005]; [0.005, 0.006] | (-0.005, -0.003]; [0.003, 0.005] | | |
| Table_Digital inquiry_Adult | II | 0 | (-∞, -0.006]; [0.006, +∞) | (-0.006, -0.005]; [0.005, 0.006] | (-0.005, -0.003]; [0.003, 0.005] | | |
| Table_Reading/learning_multifunction_Adult | I | 0 | (-∞, -0.005]; [0.005, +∞) | (-0.005, -0.004]; [0.004, 0.005] | (-0.004, -0.003]; [0.003, 0.004] | | |
| Table_Reading/learning_digital_Adult | I | 0 | (-∞, -0.005]; [0.005, +∞) | (-0.005, -0.004]; [0.004, 0.005] | (-0.004, -0.003]; [0.003, 0.004] | | |
| Table_ Hobby_Children | I | 0 | (-∞, -0.005]; [0.005, +∞) | (-0.005, -0.004]; [0.004, 0.005] | (-0.004, -0.003]; [0.003, 0.004] | | |
| Table_ Workshop_Adult | I | 0 | (-∞, -0.005]; [0.005, +∞) | (-0.005, -0.004]; [0.004, 0.005] | (-0.004, -0.003]; [0.003, 0.004] | | |

Table 4.5.1 Intervals and Context driven thresholds of SVM compared to Target (T) value.

| Visualization by colours | Health Status (HS) | CIL-I | CIL-II |
|---|---|---|---|
| | 1 | [0.2, 1] | [0.25, 1] |
| | 2 | [0.15, 0.2) | [0.2, 0.25) |
| | 3 | [0.1, 0.15) | [0.1, 0.2) |
| | 4 | [T, 0.1) | [T, 0.1) |
| | 5 | (0, T) | (0, T) |

| Objectives/thresholds | CIL | T | HS1 | HS2 | HS3 | HS4 | HS5 |
|---|---|---|---|---|---|---|---|
| Bookshelf_Display_Adult | II | 0.05 | [0.25, 1] | [0.2, 0.25) | [0.1, 0.2) | [0.05,0.1) | [0, 0.05) |
| Bookshelf_Reserve_Adult | I | 0.05 | [0.2, 1] | [0.15, 0.2) | [0.1, 0.15) | [0.05,0.1) | [0, 0.05) |
| Table_Book borrow_digital_Adult | II | 0.05 | [0.25, 1] | [0.2, 0.25) | [0.1, 0.2) | [0.05,0.1) | [0, 0.05) |
| Table_Book return_ Adult | II | 0.05 | [0.25, 1] | [0.2, 0.25) | [0.1, 0.2) | [0.05,0.1) | [0, 0.05) |
| Table Counter_multifunction_Staff | I | 0.05 | [0.2, 1] | [0.15, 0.2) | [0.1, 0.15) | [0.05,0.1) | [0, 0.05) |
| Sofa_ Reading/learning_Adult | I | 0.05 | [0.2, 1] | [0.15, 0.2) | [0.1, 0.15) | [0.05,0.1) | [0, 0.05) |
| Sofa_ Reading/learning_Children | I | 0.05 | [0.2, 1] | [0.15, 0.2) | [0.1, 0.15) | [0.05,0.1) | [0, 0.05) |
| Table_Digital inquiry_Adult | II | 0.05 | [0.25, 1] | [0.2, 0.25) | [0.1, 0.2) | [0.05,0.1) | [0, 0.05) |
| Table_Reading/learning_multifunction_Adult | I | 0.05 | [0.2, 1] | [0.15, 0.2) | [0.1, 0.15) | [0.05,0.1) | [0, 0.05) |
| Table_Reading/learning_digital_Adult | I | 0.05 | [0.2, 1] | [0.15, 0.2) | [0.1, 0.15) | [0.05,0.1) | [0, 0.05) |
| Table_ Hobby_Children | I | 0.05 | [0.2, 1] | [0.15, 0.2) | [0.1, 0.15) | [0.05,0.1) | [0, 0.05) |
| Table_ Workshop_Adult | I | 0.05 | [0.2, 1] | [0.15, 0.2) | [0.1, 0.15) | [0.05,0.1) | [0, 0.05) |

Table 4.7.1 Intervals and Context driven thresholds of EML

| Visualization by colours | Health Status (HS) | CIL-I | CIL-II |
|---|---|---|---|
| | 1 | <120 EML | <108 EML |
| | 2 | 120-132 EML | 108-120 EML |
| | 3 | 132-150 EML | 120-150 EML |
| | 4 | 150-275 EML (Standard) | 150-180 EML (Enhanced Daylight) |
| | 5 | >275 EML | >180 EML |

| Objectives/thresholds | CIL | S (EML) | HS1 | HS2 | HS3 | HS4 | HS5 |
|---|---|---|---|---|---|---|---|
| Bookshelf_Display_Adult | II | 150 | (0, 108) | [108, 120) | [120, 150) | [150, 180] | (180, +∞) |
| Bookshelf_Reserve_Adult | I | 150 | (0, 120) | [120, 132) | [132, 150) | [150, 275] | (275, +∞) |
| Table_Book borrow_digital_Adult | II | 150 | (0, 108) | [108, 120) | [120, 150) | [150, 180] | (180, +∞) |
| Table_Book return_ Adult | II | 150 | (0, 108) | [108, 120) | [120, 150) | [150, 180] | (180, +∞) |
| Table Counter_multifunction_Staff | I | 150 | (0, 120) | [120, 132) | [132, 150) | [150, 275] | (275, +∞) |
| Sofa_ Reading/learning_Adult | I | 150 | (0, 120) | [120, 132) | [132, 150) | [150, 275] | (275, +∞) |
| Sofa_ Reading/learning_Children | I | 150 | (0, 120) | [120, 132) | [132, 150) | [150, 275] | (275, +∞) |
| Table_Digital inquiry_Adult | II | 150 | (0, 108) | [108, 120) | [120, 150) | [150, 180] | (180, +∞) |
| Table_Reading/learning_multifunction_Adult | I | 150 | (0, 120) | [120, 132) | [132, 150) | [150, 275] | (275, +∞) |
| Table_Reading/learning_digital_Adult | I | 150 | (0, 120) | [120, 132) | [132, 150) | [150, 275] | (275, +∞) |
| Table_ Hobby_Children | I | 150 | (0, 120) | [120, 132) | [132, 150) | [150, 275] | (275, +∞) |
| Table_ Workshop_Adult | I | 150 | (0, 120) | [120, 132) | [132, 150) | [150, 275] | (275, +∞) |

## 5. Practical Application of the CIL Framework in Public Libraries

Building on the insights and the novel approach from previous chapters, this chapter explores the practical implementation of the CIL (Critical Integrated Level) framework through detailed case studies of three Swedish libraries, designated as G, S, and C. These studies, conducted with measurements taken in February for Library G, and in March and October 2023 for Libraries S and C respectively, utilize the GL SPECTIS 1.0 touch as the measurement tool (Fig.2).

We have chosen nine key parameters to illustrate the application of the CIL framework, categorized into visual and non-visual/semi-visual groups. The visual parameters include Illuminance ($\bar{E}_m$ in Lux), Colour Rendering Indices$R_a, R_9$, $R_f, R_g$, SDCM. The non-visual parameters assessed are Correlated Colour Temperature (CCT), Delta Ultraviolet (DUV), Stroboscopic Visibility Measure (SVM), and Equivalent Melanopic Lux (EML).

To contextualize our findings, we use the health status scale from HS1 to HS5, where HS1 denotes the worst condition and HS5 represents an extra good condition, as established in Chapter 4. Main findings from the case studies include:

- Illuminance (Lux):
  - Approximately 50% of the measurements across the libraries fall into HS1 or HS2, indicating illuminance levels significantly below the standard and target values, depicted in red.
  - Instances of excessively high illuminance were observed, hinting at potential energy inefficiencies (highlighted in dark green, HS5), which are further discussed in Chapter 6.
  - Illuminance varies significantly with shelf height, with higher shelves typically experiencing better light levels (HS5), though medium and lower shelves often remain in poor conditions (HS1).
- Colour Rendering:
  - Most data points scored well (HS4) or best (HS5) for colour rendering indices, except for notable issues with R9, particularly in Library C during conditions of low snow reflectance (October measurements), where R9 significantly worsened.
- Correlated Color Temperature (CCT):
  - CCT values exhibited dependency on the measurement's vertical position, with about 50% rating as just acceptable or poor (HS2 or HS3), potentially due to inappropriate selection or degradation of light sources.
- Non-Visual Parameters:
  - DUV: The majority of DUV measurements are classified as good (HS4) or acceptable (HS3).
  - SVM: Findings for SVM predominantly indicate good (HS4) or extra good conditions (HS5).
  - EML: EML calculations, heavily dependent on Lux and CCT, generally revealed less satisfactory statuses, warranting further discussion in Chapter 6.

This chapter not only demonstrates the practical applications of the CIL framework but also highlights the complexities and challenges encountered in achieving optimal lighting conditions in library environments. The varying results across different parameters and conditions underscore the need for tailored lighting strategies that consider both visual comfort and energy efficiency.

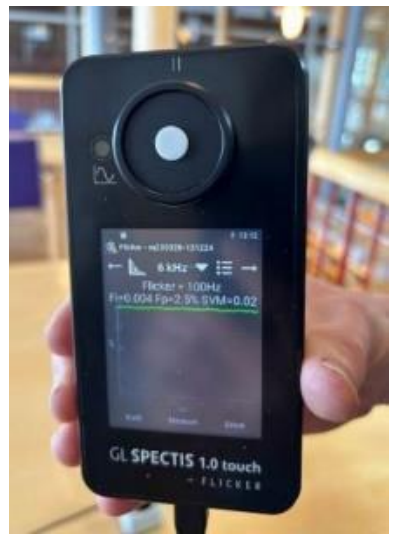

Fig.2 Measurement tool: GL SPECTIS 1.0 touch

Table 5.1 Measurement and context driven assessment at Library G_Feb2023

| Objectives/thresholds | Comments | | | CIL | Visual | | | | | | Non-Visual/semi-Visual | | |
|---|---|---|---|---|---|---|---|---|---|---|---|---|---|
| | Shelf lighting | Daylight | Position | | Lux | $R_a$ | $R_9$ | $R_f$ | $R_g$ | CCT | DUV | SVM | EML |
| Bookshelf_Display_Adult | Yes | No | High | II | 1221 | 85 | 6 | 82 | 101 | 2929 | 0.0017 | 0 | 353 |
| | | | Medium | | 344 | 86 | 14 | 86 | 98 | 2980 | 0.0008 | 0 | 105 |
| | | | Low | | 233 | 86 | 17 | 87 | 98 | 3040 | 0.0015 | 0 | 72 |
| Bookshelf_Display_Adult | No | Yes | High | | 240 | 95 | 61 | 95 | 98 | 4088 | 0.0026 | 0 | 98 |
| | | | Medium | | 265 | 95 | 63 | 95 | 98 | 4450 | 0.0043 | 0 | 114 |
| | | | Low | | 270 | 94 | 57 | 95 | 98 | 4150 | 0.0021 | 0 | 110 |
| Bookshelf_Display_Adult | No | No | High | | 125 | 86 | 20 | 86 | 100 | 2396 | -0.0042 | 0 | 34 |
| | | | Medium | | 284 | 86 | 19 | 87 | 98 | 2755 | -0.00224 | 0 | 84 |
| | | | Low | | 320 | 85 | 16 | 86 | 97 | 2828 | -0.0012 | 0 | 96 |
| Table Counter_multifunction_Staff | / | / | / | I | 370 | 84 | 14 | 86 | 96 | 3060 | 0.0005 | 0 | 116 |
| Sofa_ Reading/learning_Children | / | No | / | I | 700 | 86 | 13 | 86 | 99 | 3040 | 0.0008 | 0 | 215 |
| Table_Reading/learning_multifunction_Adult | / | Yes | / | I | 770 | 87 | 26 | 88 | 97 | 3360 | -0.0003 | 0 | 263 |
| Table_Reading/learning_digital_Adult | / | / | Table | I | 470 | 84 | 14 | 86 | 96 | 3070 | 0.0003 | 0 | 147 |
| | / | / | Screen | | 260 | 88 | 29 | 89 | 97 | 3250 | 0.0004 | 0 | 87 |
| Table_ Hobby_Children | / | No | / | I | 447 | 83 | -1* | 79 | 101 | 2838 | 0.0049 | 0 | 122 |

*: abnormal value(s)

Table 5.2 Measurement and context driven assessment at Library S_Mar2023

| Objectives/thresholds | Comments | | | CIL | Visual | | | | | | Non-Visual/semi-Visual | | |
|---|---|---|---|---|---|---|---|---|---|---|---|---|---|
| | Shelf lighting | Daylight | Position | | Lux | $R_a$ | $R_9$ | $R_f$ | $R_g$ | CCT | DUV | SVM | EML |
| Bookshelf_Display_Adult | No | No | High | II | 740 | 89 | 40 | 89 | 97 | 3010 | -0.0019 | 1 | 241 |
| | | | Medium | | 279 | 94 | 63 | 92 | 98 | 2911 | -0.0016 | 0 | 92 |
| | | | Low | | 268 | 94 | 63 | 92 | 98 | 2902 | -0.0016 | 0 | 88 |
| Bookshelf_Display_Adult | No | No | High | | 389 | 94 | 70 | 92 | 99 | 2675 | -0.0031 | 4* | 123 |
| | | | Medium | | 185 | 95 | 71 | 92 | 99 | 2740 | -0.0032 | 0 | 59 |
| | | | Low | | 139 | 95 | 72 | 92 | 100 | 2734 | -0.0033 | 0 | 44 |
| Bookshelf_Display_Adult | No | Yes | High | | 1154 | 94 | 65 | 93 | 99 | 3411 | 0.0005 | 0 | 420 |
| | | | Medium | | 215 | 95 | 73 | 95 | 98 | 3670 | 0.0022 | 0 | 83 |
| | | | Low | | 215 | 95 | 70 | 95 | 98 | 3480 | 13* | 0 | 80 |
| Bookshelf_Display_Adult/Children | No | No | Medium | | 115 | 93 | 63 | 93 | 99 | 3071 | 0.0003 | 0 | 39 |
| | | | Low | | 113 | 94 | 67 | 94 | 100 | 3112 | 0.0003 | 0 | 39 |
| Table Counter_multifunction_Staff | No | No | / | I | 300 | 94 | 64 | 93 | 100 | 3219 | -0.0006 | 0 | 135 |
| Sofa_ Reading/learning_Adult | No | Yes | / | I | 1208 | 93 | 61 | 93 | 99 | 3250 | 0.0011 | 0 | 424 |
| Table_Reading/learning_multifunction_Adult | No | No | Table | I | 735 | 95 | 71 | 94 | 99 | 3602 | 0.0001 | 0 | 281 |
| | | | Screen | I | 650 | 94 | 69 | 94 | 99 | 3569 | 0.0006 | 0 | 241 |
| Table_ Hobby_Children | No | Yes | / | I | 1611 | 96 | 76 | 95 | 98 | 4056 | 0.0018 | 0 | 667 |

Table 5.3 Measurement and context driven assessment at Library S_Oct2023

| Objectives/thresholds | Comments | | | CIL | Visual | | | | | | Non-Visual/semi-Visual | | |
|---|---|---|---|---|---|---|---|---|---|---|---|---|---|
| | Shelf lighting | Daylight | Position | | Lux | $R_a$ | $R_9$ | $R_f$ | $R_g$ | CCT | DUV | SVM | EML |
| Bookshelf_Display_Adult | No | No | High | II | 570 | 90 | 44 | 89 | 97 | 2980 | -0.0019 | 0 | 185 |
| | | | Medium | | 300 | 93 | 63 | 91 | 98 | 2900 | -0.0016 | 0 | 98 |
| | | | Low | | 210 | 93 | 62 | 91 | 98 | 2900 | -0.0016 | 0 | 69 |
| Bookshelf_Display_Adult | No | No | High | | 250 | 94 | 72 | 92 | 99 | 2640 | -0.0034 | 0 | 79 |
| | | | Medium | | 122 | 95 | 72 | 92 | 100 | 2620 | -0.0035 | 0 | 38 |
| | | | Low | | 83 | 95 | 73 | 93 | 100 | 2600 | -0.0037 | 0 | 28 |
| Bookshelf_Display_Adult | No | Yes | High | | 1100 | 93 | 63 | 93 | 99 | 3300 | 0.0007 | 0 | 390 |
| | | | Medium | | 300 | 93 | 62 | 92 | 99 | 3000 | 0.0004 | 0 | 100 |
| | | | Low | | 230 | 93 | 62 | 92 | 99 | 3050 | 0.0004 | 0 | 104 |
| Bookshelf_Display_Adult/Children | No | No | Medium | | 100 | 96 | 73 | 95 | 99 | 3250 | -0.0004 | 0 | 36 |
| | | | Low | | 74 | 94 | 65 | 93 | 100 | 2950 | -0.0006 | 0 | 24 |
| Table Counter_multifunction_Staff | No | No | / | I | 340 | 93 | 60 | 92 | 99 | 3200 | -0.0002 | 0 | 130 |
| Sofa_ Reading/learning_Adult | No | Yes | / | I | 1070 | 94 | 66 | 93 | 100 | 3300 | -0.0002 | 0 | 379 |
| Table_Reading/learning_multifunction_Adult | No | No | Table | I | 160 | 93 | 62 | 93 | 99 | 2900 | -0.0015 | 0 | 52 |
| | | | Screen | | 225 | 96 | 71 | 95 | 99 | 3560 | -0.0005 | 0 | 86 |
| Table_ Hobby_Children | No | Yes | / | I | 1150 | 94 | 68 | 94 | 98 | 3700 | 0.0009 | 0 | 443 |

Table 5.4 Measurement and context driven assessment at Library C_Mar2023

| Objectives/thresholds | Comments | | | CIL | Visual | | | | | | Non-Visual/semi-Visual | | |
|---|---|---|---|---|---|---|---|---|---|---|---|---|---|
| | Shelf lighting | Daylight | Position | | Lux | $R_a$ | $R_9$ | $R_f$ | $R_g$ | CCT | DUV | SVM | EML |
| Bookshelf_Display_Adult | Yes | Yes | High | II | 1390 | 92 | 37 | 89 | 100 | 3442 | -0.0005 | 0.01 | 486 |
| | | | Medium | | 820 | 95 | 58 | 94 | 100 | 3800 | 0.0004 | 0 | 321 |
| | | | Low | | 800 | 97 | 69 | 96 | 100 | 4200 | 0.0007 | 0 | 358 |
| Bookshelf_Display_Adult | No | Yes | High | | 730 | 97 | 98 | 98 | 101 | 5657 | -0.0002 | 0 | 360 |
| | | | Medium | | 335 | 95 | 69 | 95 | 101 | 3558 | -0.0027 | 0 | 128 |
| | | | Low | | 177 | 94 | 73 | 94 | 102 | 3431 | -0.0043 | 0 | 67 |
| Table Counter_multifunction_Staff | Yes | No | / | I | 506 | 95 | 65 | 95 | 99 | 3914 | 0.0016 | 0 | 200 |
| Table_Reading/learning_multifunction_Adult | No | No | Tabel | I | 675 | 97 | 80 | 97 | 98 | 4758 | 0.0053 | 0 | 309 |
| | | | Screen | | 380 | 98 | 90 | 98 | 99 | 4458 | 0.0021 | 0 | 170 |
| Table_Reading/learning_digital_Adult | No | Yes | / | I | 1100 | 97 | 76 | 95 | 101 | 4747 | -0.0009 | 0 | 488 |
| Table_ Hobby_Children | No | No | / | I | 628 | 98 | 82 | 97 | 99 | 4258 | 0.0025 | 0 | 270 |
| Table_ Workshop_Adult | Yes | Yes | / | I | 770 | 91 | 33 | 88 | 101 | 3340 | -0.0012 | 0 | 264 |

Table 5.5 Measurement and context driven assessment at Library C_Oct2023

| Objectives/thresholds | Comments | | | CIL | Visual | | | | | | Non-Visual/semi-Visual | | |
|---|---|---|---|---|---|---|---|---|---|---|---|---|---|
| | Shelf lighting | Daylight | Position | | Lux | $R_a$ | $R_9$ | $R_f$ | $R_g$ | CCT | DUV | SVM | EML |
| Bookshelf_Display_Adult | Yes | Yes | High | II | 970 | 85 | 3 | 80 | 102 | 2900 | 0.0008 | 0 | 276 |
| | | | Medium | | 400 | 88 | 11 | 82 | 102 | 2770 | -0.0005 | 0 | 115 |
| | | | Low | | 310 | 89 | 11 | 85 | 102 | 2700 | -0.0011 | 0 | 91 |
| Bookshelf_Display_Adult | No | Yes | High | | 110 | 94 | 59 | 93 | 101 | 3400 | -0.0017 | 0 | 40 |
| | | | Medium | | 130 | 95 | 58 | 93 | 101 | 3500 | -0.0013 | 0 | 48 |
| | | | Low | | 80 | 95 | 58 | 93 | 101 | 3500 | -0.0013 | 0 | 30 |
| Table Counter_multifunction_Staff | Yes | No | / | I | 570 | 87 | 23 | 85 | 101 | 3100 | 0.002 | 0 | 177 |
| Table_Reading/learning_multifunction_Adult | No | No | Tabel | I | 190 | 96 | 64 | 96 | 99 | 3900 | 0.0015 | 0 | 76 |
| | | | Screen | | 100 | 94 | 54 | 93 | 101 | 3300 | -00009 | 0 | 35 |
| Table_Reading/learning_digital_Adult | No | Yes | / | I | 750 | 96 | 68 | 95 | 100 | 4400 | 0.0016 | 0 | 320 |
| Table_ Hobby_Children | No | No | / | I | / | / | / | / | / | / | / | / | / |
| Table_ Workshop_Adult | Yes | Yes | / | I | 600 | 90 | 35 | 87 | 101 | 3400 | 0.0004 | 0 | 205 |

## 6. Discussions

Building upon the case studies detailed in Chapter 5, this chapter delves into a discussion of the principal findings from these studies. Our focus centres on the degradation of key lighting parameters, the nuances of Equivalent Melanopic Lux (EML), a simplified comparison on measurement results from different seasons and different libraries, and the intricacies of energy efficiency. Each of these elements plays a pivotal role in shaping the efficacy and sustainability of lighting asset management in public libraries.

### 6.1 Key lighting parameters degradation issues

Our analysis has revealed critical areas where lighting performance within public libraries needs urgent enhancement, specifically concerning Lux, $R_9$, CCT, and DUV see Table 5.1 – Table5.5. The degradation of these parameters over time significantly impacts the overall effectiveness of lighting systems. Moreover, the calculations of Equivalent Melanopic Lux (EML) are directly influenced by Lux and CCT values, suggesting that their degradation adversely affects EML outcomes.

Under the assumption that the initial design and installation being adequate, then the status of assessing those key lighting parameters shows that Lux, $R_9$, CCT, and DUV deteriorate with time. This natural progression underscores the necessity of incorporating these parameters into lifetime predictions and optimizing operational and maintenance strategies to extend the effective lifespan of lighting installations.

Current standards for defining the reliability of lighting sources, particularly LEDs, are predominantly cantered on the" L70" benchmark [21]. This metric indicates when the luminous flux of LEDs declines to 70% of its initial output, traditionally used to estimate the lifespan of LED lighting. However, our findings advocate for a broader approach. Considering the degradation trends of not just Lux but also $R_9$ , CCT and DUV is crucial as these factors are integral to maintaining optimal lighting conditions and enhancing the EML values.

Additionally, the vertical position's impact on degradation is notable, with varying degrees of deterioration observed at different heights within the library spaces. This variation suggests that the commonly specified measurement height of 0.75 meters [14] may be insufficient to capture the full scope of spatial lighting behaviour and degradation trends. Expanding the range of measurement heights could provide a more comprehensive understanding of lighting performance across different levels of a library's environment.

Overall, addressing these degradation issues not only involves revisiting maintenance schedules and operational practices but also calls for a re-evaluation of the metrics used to define lighting reliability and effectiveness. By broadening the scope of these metrics to include $R_9$, CCT and DUV alongside Lux, and considering the influence of vertical positioning, library lighting systems can be better maintained to serve their purpose over longer periods, ensuring both energy efficiency and high-quality illumination.

### 6.2 Circadian Rhythms and EML

Daylight emerges as the most effective source of melanopic lux, as evidenced by its spectral composition and energy efficiency in comparison to artificial lighting sources. Our testing revealed that the presence of daylight yielded superior performance across various metrics in both visual and non-visual/semi-visual domains. Notably, daylight serves as a potent circadian stimulus, playing a crucial role in regulating circadian rhythms, influencing sleep-wake cycles, hormone production, and overall well-being.

While the integration of daylight into architectural design is commendable and reduces reliance on electrical lighting, the significant variability of sunlight throughout the year poses challenges. Fig. 3 illustrates this variability, ranging from summer's midnight sun to winter's prolonged darkness. Such fluctuations present difficulties in fenestration placement, shading system control, and management of artificial lighting assets. Particularly during the highlighted orange timeframe of the year, the prolonged influence of natural lighting dynamics on human photoreceptors is evident. Exposure to natural daylight benefits cone cells, sensitive to bright light and colour perception, as well as ganglion cells, sensitive to blue light, thereby promoting alertness during daytime hours [22].

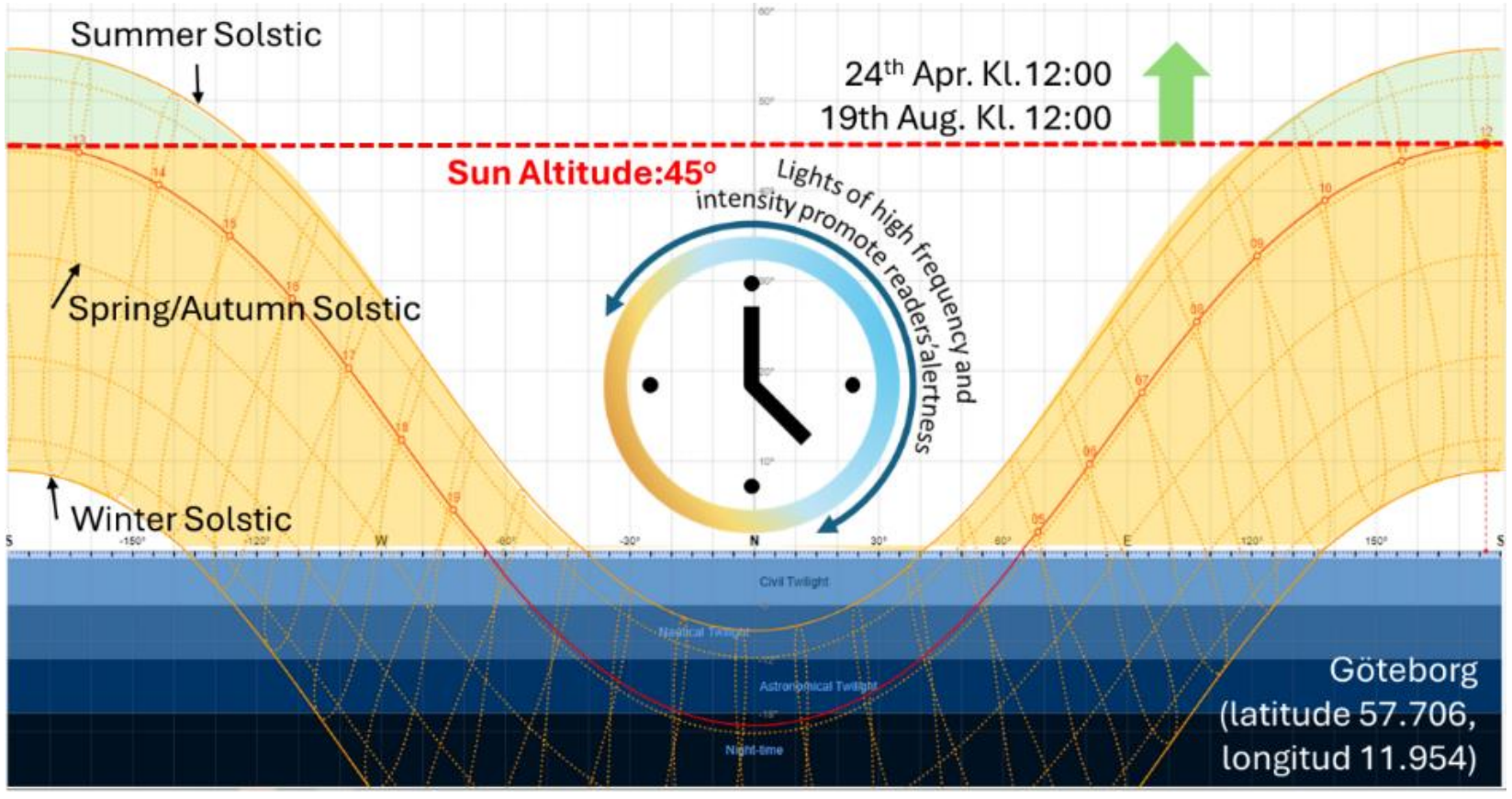

Fig.2 Cartesian sun-path diagram with recommended healthy CCT scale of Gothenburg, Sweden (Note: In our case study, Library G is in the city of Gothenburg)

### 6.3 Comparison studies: from different seasons and different libraries

This section utilizes the median function to analyse metrics across three measured categories—visual parameters (Lux, $R_9$, and CCT), and non-visual parameters (DUV, SVM, and EML). Based on the results from Table 5.1-5.5, in this section, the Health Status (HS) of parameters is scored on a scale of 1 to 5 for each measurement, with each HS level corresponding to a specific score as detailed in Table 6.1. The median scores for all parameters are then compiled and presented in the final row of the table.

Table 6.1 From HS to Scores

| Objectives/thresholds | Position | Visual | | | Non-Visual/semi-Visual | | |
|---|---|---|---|---|---|---|---|
| | | Lux | $R_9$ | CCT | DUV | SVM | EML |
| Bookshelf_Display_Adult | High | 5 | 1 | 3 | 1 | 5 | 5 |
| | Medium | 3 | 1 | 3 | 5 | 5 | 1 |
| | Low | 1 | 1 | 3 | 4 | 5 | 1 |
| Bookshelf_Display_Adult | High | 1 | 5 | 5 | 4 | 5 | 1 |
| | Medium | 1 | 5 | 5 | 3 | 5 | 2 |
| | Low | 2 | 4 | 5 | 4 | 5 | 2 |
| Bookshelf_Display_Adult | High | 1 | 1 | 2 | 3 | 5 | 1 |
| | Medium | 2 | 1 | 2 | 4 | 5 | 1 |
| | Low | 3 | 1 | 3 | 4 | 5 | 1 |
| Table Counter_multifunction_Staff | / | 1 | 1 | 2 | 5 | 5 | 1 |
| Sofa_ Reading/learning_Children | / | 4 | 1 | 4 | 5 | 5 | 4 |
| Table_Reading/learning_multifunction_Adult | / | 3 | 1 | 2 | 5 | 5 | 4 |
| Table_Reading/learning_digital_Adult | Table | 1 | 1 | 2 | 5 | 5 | 2 |
| | Screen | 1 | 1 | 2 | 5 | 5 | 1 |
| Table_ Hobby_Children | / | 1 | 1 | 2 | 2 | 5 | 2 |
| Median | | 1 | 1 | 3 | 4 | 5 | 1 |

Following the outlined approach, we have compiled scores for all five measurements in Table 6.2. The average scores of these five measurements, representing a single data set or group, are displayed in the final row of the table.

Table 6.2 Scores of each library and average values

| Libraries/Average | Lux | $R_9$ | CCT | DUV | SVM | EML |
|---|---|---|---|---|---|---|
| Library G_Feb2023 | 1 | 1 | 3 | 4 | 5 | 1 |
| Library S_Mar2023 | 1 | 5 | 3 | 4 | 5 | 2 |
| Library S_Oct2023 | 1 | 5 | 3 | 5 | 5 | 1 |
| Library C_Mar2023 | 4 | 5 | 5 | 4 | 5 | 4 |
| Library C_Oct2023 | 1 | 3 | 3 | 4 | 5 | 1 |
| Library_Average | 1.6 | 3.8 | 3.4 | 4.2 | 5 | 1.8 |

As detailed in Table 6.2, the results are depicted in Figure 3, which presents data from Library S and Library C for the months of March and October, represented in blue and orange to signify spring and autumn, respectively. The average scores from these measurements are also highlighted in green within the figures.

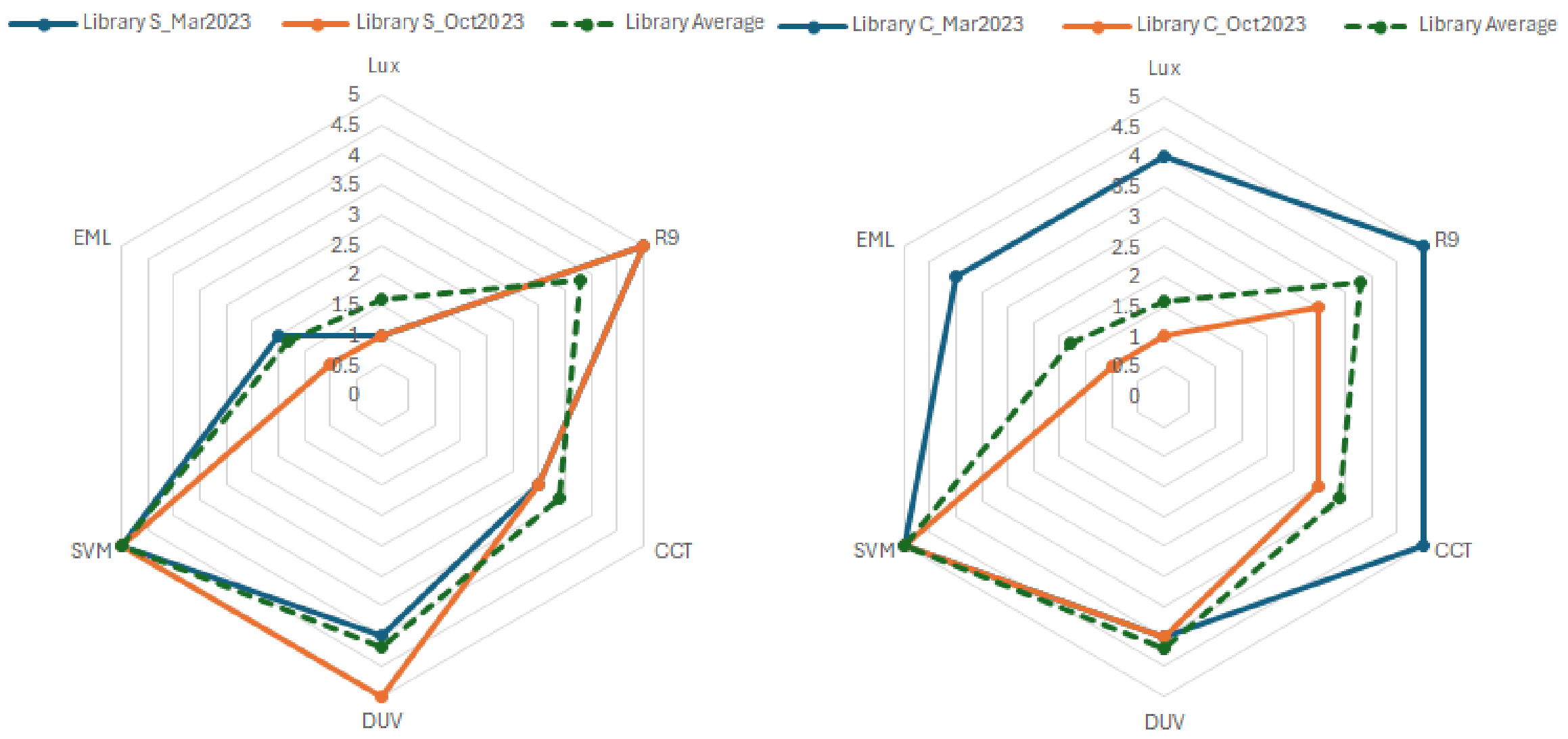

(a) Library S (b) Library C

Fig.3 Radar diagram with median values from the measured visual, non-visual categories

The comparative analysis highlights distinct seasonal variations between the two libraries. While the differences in measurements between spring and autumn are minimal for Library S, they are more pronounced in Library C. This variance can largely be attributed to the architectural features of Library C, which includes extensive glass facades that significantly increase the influence of snow reflection during March. This finding underscores how human-centric and integrative lighting asset management strategies can be significantly affected by the physical characteristics of the building, particularly the presence of large glass areas (as also noted in reference [6]). This seasonal comparison not only reflects the direct impact of architectural elements on lighting conditions but also emphasizes the need for adaptive lighting strategies that can accommodate such variations effectively.

The comparative analysis highlights variations among the three libraries. Observing the green lines, which represent the average measurements, we note distinct differences: For Library S, the lighting performance is close to the average. In contrast, Library C performs above average, although it dips below the average in October. This comparison against average levels underscores the disparities among libraries. Incorporating more measurements into the analysis provides deeper insights, enabling library asset managers to make more informed decisions tailored to specific needs at various levels.

### 6.4 Insights on Energy Efficiency

Typically, the energy efficiency of lighting sources is determined by dividing the declared useful luminous flux (in Lumens) by the declared on-mode power consumption (in Watts) and then multiplying by the applicable factor, resulting in ratings from A to G [11]. However, this method does not fully account for lighting performance or human needs in real indoor environments.

In terms of Lighting Asset Management, it is crucial to achieve an optimal balance between adequate daylighting and the mitigation of solar heat load to ensure an optimal indoor environment. For instance, during sunny winter periods, the use of shades to block excessive sunlight while still relying on artificial lighting can lead to inefficient energy use. Similarly, in sunny summer conditions, excessive solar gain can compromise thermal comfort, thereby increasing the need for cooling, which in turn raises energy consumption. These observations are consistent with the qualitative insights from our studies [6], which are further supported by quantitative measurements presented in Chapter 5.

Swedish sustainable building certification standards, such as Miljöbyggnad [23], address these issues by incorporating the Solar Heat Load matrix, which quantifies solar heat transmitted through windows per square meter of floor area. This approach provides facility managers in libraries with a holistic view of fenestration strategies, shading control, and shading types.

Our case studies suggest that the energy efficiency of human-centric and integrated lighting assets should be evaluated based not only on energy consumption and individual lighting source' performance, but also on integrated lighting performance parameters in the real environment, such as Lux, $R_9$, CCT, DUV, SVM, and EML. Once the types of lighting sources in a library are established, their energy consumption can be considered constant. However, as the performance of these lighting parameters degrades over time, so too does the energy efficiency of the lighting assets. This degradation implies that energy efficiency diminishes over time and can be predicted based on the integrated degradation of these lighting parameters.

## 7. Conclusions and Future Research

To address the challenge of managing human-centric and integrative lighting in public libraries, particularly post-installation, it is crucial to balance complex visual and non-visual light effects. This study introduces a novel, context-driven Critical Integrative Levels (CIL) approach to tackle these issues. The main contributions from this study include:

- **Identification of Key Parameters and Contexts:** We identified critical parameters and contexts that significantly influence lighting asset management in public libraries, enhancing the understanding of what factors are most impactful.
- **Introduction of MTOE and CIL Framework:** We introduced the concept of Mean Time of Exposure (MTOE) and developed the Critical Integrative Levels (CIL) framework for managing lighting assets. This approach includes foundational assumptions and illustrative examples, providing a structured method to address the complexities of lighting management in library settings.
- **Establishment of Context-Specific Thresholds:** Utilizing the CIL framework, we established and defined context-specific thresholds for essential lighting parameters. This development allows for targeted and effective management strategies tailored to the unique needs of public libraries.

- **Practical Application of the CIL Framework:** The efficacy of the CIL approach was demonstrated through practical applications in real-world settings. Case studies from three Swedish libraries showcased how the selected parameters could be optimized to enhance both functionality and comfort.
- **Promotion of Integrative Dialogue:** Our study encourages ongoing dialogue regarding the integration of Equivalent Melanopic Lux (EML) with energy efficiency solutions. This conversation aims to foster a holistic approach to lighting design that prioritizes human well-being alongside environmental sustainability.

The study's findings are pivotal in setting a new precedent for lighting asset management in public libraries. However, the research is not without its limitations, which include the constraints imposed by the selected parameters and the foundational assumptions of the CIL framework. The following areas are identified for further development:

- **Integrated Prognosis Approach:** Future work will explore the development of an integrated prognosis approach that combines the CIL framework with advanced digital solutions. This progression will aim to enhance predictive capabilities and improve maintenance strategies, making them more proactive and data driven.
- **Digital Integration:** The potential for incorporating digital solutions such as Digital Twins and IoT-based monitoring systems will be examined. These technologies promise to revolutionize how lighting assets are managed, monitored, and maintained over their lifecycle, ensuring optimal performance and minimal environmental impact.
- **Expansion of Parameters and Contexts:** Further research will also involve expanding the range of parameters and contexts considered within the CIL framework to cover more diverse and comprehensive scenarios. This expansion will likely include more nuanced aspects of human-centric and integrated lighting and broader environmental considerations.

## Acknowledgments

We extend our gratitude to the Swedish Energy Agency (Energimyndigheten) for their financial support, which made this research possible. The project, "Integrated Lighting Asset Management in Public Libraries (Integrerad tillgångsförvaltning för belysning i allmänna bibliotek genom Digital Tvilling)", bearing the project number P2022-00277, has benefited immensely from their backing.

We are profoundly grateful to Dr. Christofer Silfvenius from Scania AB, the visionary initiator behind the project idea. His original concept and ongoing support have been pivotal in steering the direction and focus of our study. Special thanks are due to Dr. Jörgen Sjödin, the project manager at the Swedish Energy Agency, whose expert guidance has been indispensable throughout the research process.

We appreciate the insights on the parameters from discussion with Klas Rejgård from Lighting Standard Specialist Fagerhults.

Lastly, we owe a great debt of gratitude to all the participants from the 20 public libraries, whose active involvement was crucial. The managers, staff, and regular users generously contributed their time and shared their experiences, offering a wealth of insights that have been central to our understanding of integrated lighting asset management in public libraries. Their input has been invaluable, and we are truly appreciative of their collaboration.

Table.2.1 key lighting parameters on its performance measurement

| Category | Sub-Category | Comments |
|---|---|---|
| Illuminance | Average Illuminance | This refers to the average level of light distributed over a particular surface or area. It is generally used to ensure sufficient lighting for tasks carried out in that area. |
| | Maximum Illuminance | This refers to the point of highest light level on a given surface. It's important to avoid excessive maximum illuminance that can cause glare and discomfort. |
| | Minimum Illuminance | This refers to the point of lowest light level on a given surface. It's crucial to ensure the minimum illuminance is still sufficient for visual tasks to prevent strain and fatigue. |
| | Uniformity Ratio of Illuminance | This is the ratio of minimum to average or maximum illuminance. Uniformity is key to avoid sharp contrasts in lighting, which can cause visual discomfort and impair task performance. |
| | Maintained Illuminance | This is the illuminance level that lighting installations are designed to provide over most of their operational life, considering factors such as depreciation of lamp output and room surface dirt depreciation. |
| Glare | Direct Glare | This occurs when bright lights are directly in the field of view. It is usually caused by inadequately shielded light sources and can cause discomfort and reduced visual performance. |
| | Discomfort Glare | This type of glare causes discomfort without necessarily impairing the clarity of a scene. It is often caused by high contrasts or extreme brightness levels within the field of view. |
| | Unified Glare Rating (UGR) | UGR is a numerical measure of the glare in a particular environment, considering factors such as the size, position, and luminance of light sources, as well as the viewing angle. Lower UGR values correspond to less glare. |
| | Glare Rating (GR) | This is another numerical measure of glare, primarily used in outdoor lighting. Like UGR, lower GR values indicate less glare. |
| | Glare by Reflection | This occurs when light sources are reflected off shiny or glossy surfaces, causing bright spots that can distract or impair vision. |
| | Veiling Reflection | This is a type of glare caused by reflections that wash out the colour s and contrast in a scene, making it difficult to see details clearly. |
| Colour rendering | Colour rendering Index (CRI) | The CRI is a quantitative measure of the ability of a light source to reveal the colour s of various objects faithfully compared to an ideal or natural light source. The index ranges from 0 to 100, with a CRI of 100 representing the maximum possible colour rendering quality. |
| | General Colour rendering Index ($R_a$) | This is the average of the first eight colour rendering indices (R1 through R8) in the CRI calculation, which represent common colour s such as light and medium skin tones, sky blue, foliage green, etc. Like CRI, it is a measure of how accurately a light source can render these colour s compared to natural light. |
| | Special Colour rendering Index ($R_i$) | This includes the colour rendering indices from R9 to R14, which represent saturated colour s such as red, yellow, blue, and green. These indices are important in situations where accurate colour rendering of these specific colour s is crucial. |
| | Fidelity Index ($R_f$) | The Fidelity Index is part of the IES TM-30 colour rendering system, which is an advanced method of evaluating colour rendition. Rf measures how closely the colour rendering of a light source match that of a reference illuminant, with values ranging from 0 to 100. |
| | Gamut Index ($R_g$) | Also part of the IES TM-30 system, the Gamut Index measures the average level of colour saturation when illuminated by the test source compared to the reference illuminant. |
| | Standard Deviation of Colour Matching (SDCM) | The term "MacAdam Ellipse" or "MacAdam Step" originates from David Lewis MacAdam's 1942 research, which established that the human eye cannot detect color differences within specific ellipses in chromaticity space. These ellipses measure color deviation with a unit called the "Standard Deviation of Colour Matching" (SDCM), also known as "MacAdam SDCM." One SDCM, or one MacAdam Step, represents the smallest noticeable deviation. A color tolerance of 3 SDCM, for instance, indicates that color variations are within three MacAdam ellipses, and such differences are typically imperceptible. This metric is crucial in lighting design and quality control, particularly for ensuring color consistency across LED lighting, where color output can vary significantly between units. |
| Colour Temperature | Correlated Colour temperature (CCT) | CCT is a specification of the colour appearance of the light emitted by a lamp, relating its colour to the colour of light from a reference source when heated to a particular temperature, measured in degrees Kelvin (K). Lower CCTs (2000-3000K) are described as 'warm' and have a reddish-yellowish appearance, while higher CCTs (5000-6500K) are referred to as 'cool' and have a bluish-white appearance. |
| | Duv | This is a measure of the deviation of a white light source from the black-body line in the CIE 1960 colour space. A Duv of zero means the colour point falls on the black-body line, indicating that the light source has the purest white colour corresponding to its CCT. A positive Duv indicates a shift towards green, while a negative Duv indicates a shift towards magenta. The DUV value that is considered acceptable for human perception and application requirements can vary depending on the context and the specific standards applied. In general, for most indoor lighting applications where color quality is important, such as in residential or office settings, a DUV value close to zero is preferred, typically within the range of -0.003 to +0.003. |
| Flicker | Flicker Frequency | This is the rate at which the light output fluctuates. It's measured in Hertz (Hz). A higher frequency is generally less noticeable and thus less likely to cause discomfort or health issues. |
| | Flicker Percentage | This measures the amount of fluctuation in light output, usually as a percentage of the total light output. Higher percentages generally mean the flicker is more noticeable. |
| | Flicker Index | This is a more complex measure that considers both the amplitude and the waveform of the flicker. It ranges from 0 (no flicker) to 1 (maximum flicker). |
| | SVM | SVM (Stroboscopic Visibility Measure) measures the visibility of stroboscopic effects from temporal light modulation, which can occur when moving objects are illuminated by a flickering light source. |
| Chromaticity | Chromaticity Coordinates | In a chromaticity diagram, the colour of the light is represented by a point defined by two (in CIE 1931 or CIE 1960 space) or three (in CIE 1976 space) chromaticity coordinates. The most used are the x and y coordinates in the CIE 1931 space. For instance, the chromaticity coordinates of the D65 illuminant (representing average daylight) are (0.3127, 0.3290) in the CIE 1931 space. |
| | Chromaticity Diagram | This is a plot of the chromaticity coordinates. In the CIE 1931 chromaticity diagram, the outer edge represents the spectrum of visible light, the colour s of the rainbow from red to violet, and the interior points represent the nonspectral colour s which can be formed by mixing light of different wavelengths. |
| | Chromaticity Tolerances | This refers to the allowable deviation from a specified set of chromaticity coordinates, representing the colour consistency of a batch of lamps or luminaire. It is usually represented by a small area on the chromaticity diagram, within which the chromaticity coordinates of all samples must fall. The MacAdam ellipse is a common way to define chromaticity tolerances, where a one-step MacAdam ellipse is barely noticeable to the human eye. |
| Photobiological Safety | Ultraviolet, UV | Photobiological safety refers to the measures and standards in place to protect humans from the potential harmful effects of exposure to optical radiation, including ultraviolet (UV) light, visible light, and infrared radiation. It primarily concerns mitigating risks associated with UV exposure, which can range from minor to severe health issues. |
| Circadian Rhythm | Equivalent Melanopic Lux (EML) | Equivalent Melanopic Lux (EML) is a vital metric for evaluating light's impact on human health, particularly regarding circadian rhythms and cognitive functions. It centers on how light affects the eye's melanopsin-containing cells, which influence sleep, mood, and alertness. |
| Energy Efficiency | Luminous Efficacy | This is a measure of how well a light source produces visible light. It is calculated by dividing the luminous flux (in lumens, lm) by the power consumed to produce that light (in watts, W). The higher the luminous efficacy, the more efficient the light source is at converting electricity into visible light. |
| | Lighting Power Density (LPD) | LPD is a measure of the power used by a lighting system per unit area of the space lit by the system. It's often measured in watts per square foot (W/ft^2) or watts per square meter (W/m^2). Lower LPDs mean that the lighting system is more energy efficient. |
| | Luminous Efficiency | This is the ratio of the total luminous flux to the total radiant flux (total light output divided by total energy output). It is a measure of how efficiently a light source converts radiant energy into visible light. |
| | Energy Star Rating | In some regions, lighting products can receive an Energy Star rating if they meet certain criteria for energy efficiency. The Energy Star program is run by the U.S. Environmental Protection Agency and the U.S. Department of Energy. |
| Lifespan | Rated Life | This is the estimated lifespan of a light source, provided by the manufacturer, based on standardized testing procedures. It represents the median operational life expectancy. For example, a typical incandescent light bulb may have a rated life of 1,000 hours. |
| | L70 and L50 Life | These are measures of how long it takes for a light source to degrade to a certain percentage of its original light output. L70 life is the estimated hours of operation until the light output degrades to 70% of its initial level. Similarly, L50 life is the time until the light output degrades to 50%. These are often used for LED light sources, which don't typically "burn out" like other light sources, but gradually decrease in brightness over time. |
| | Survival Rate | This is the percentage of a batch of lamps that continue to operate after a certain number of hours. For example, if 90 out of 100 lamps are still working after 1,000 hours, the survival rate is 90%. |
| | Maintenance Factor | This factor considers the reduction in light output over time (lamp lumen depreciation), as well as other factors such as lamp failures, luminaire surface depreciation, and room surface dirt depreciation. |
| Luminaire | Shielding Angle of Luminaire | This is the angle below horizontal at which the lamp or light source in the luminaire is not directly visible, providing a measure of the luminaire's ability to control glare. |
| | Ingress Protection (IP) Rating | This is a rating system that defines the ability of a luminaire to prevent the ingress of dust and water. It's given as 'IP' followed by two digits, the first digit indicating the level of protection against solid objects and the second digit indicating the level of protection against liquids. |